\documentclass[superscriptaddress,altaffilletter,aps,pra,reprint,twocolumn,showkeys,letterpaper,amsmath,amssymb,aps,floatfix]{revtex4-1}

\usepackage{graphicx}
\usepackage{dcolumn}
\usepackage{array}
\usepackage{color}

\usepackage{amsmath}

\usepackage{bm}
\usepackage{hyperref}
\usepackage{ulem}

\newcommand{\omegaMW}{\omega_{\textrm{MW}}}
\newcommand{\omegaRF}{\omega_{\textrm{RF}}}
\newcommand{\omegaHF}{\omega_{\textrm{hfs}}}

\begin{document}

\title{Decoherence-free radiofrequency dressed subspaces}
\author{G.A. Sinuco-Leon}
\thanks{G.A.S-L and H.M contributed equally.}
\affiliation{Department of Physics \& Astronomy, University of Sussex, Falmer, Brighton, BN1 9QH, UK}

\author{H. Mas}
\altaffiliation[Current affiliation: ]{Propulsion Laboratory, California Institute of Technology, Pasadena, California 91109, USA}
\thanks{G.A.S-L and H.M contributed equally.}
\affiliation{Institute of Electronic Structure and Laser, Foundation for Research and Technology-Hellas, Heraklion 70013, Greece}
\affiliation{ Department of Physics, University of Crete, Heraklion, Greece}

\author{S. Pandey}
\altaffiliation[Current affiliation: ]{Los Alamos National Laboratory, Los Alamos, NM 87545, USA}
\affiliation{Institute of Electronic Structure and Laser, Foundation for Research and Technology-Hellas, Heraklion 70013, Greece}
\affiliation{Department of Materials Science and Technology, University of Crete, Heraklion, Greece}

\author{G. Vasilakis}
\affiliation{Institute of Electronic Structure and Laser, Foundation for Research and Technology-Hellas, Heraklion 70013, Greece}

\author{B.M. Garraway}
\affiliation{Department of Physics \& Astronomy, University of Sussex, Falmer, Brighton, BN1 9QH, UK}

\author{W. von Klitzing}
\affiliation{Institute of Electronic Structure and Laser, Foundation for Research and Technology-Hellas, Heraklion 70013, Greece}

\date{\today}

\begin{abstract}
We study the spectral signatures and coherence properties of radiofrequency dressed hyperfine Zeeman sub-levels of $^{87}$Rb. Experimentally, we engineer combinations of static and RF magnetic fields to modify the response of the atomic spin states to environmental magnetic field noise. We demonstrate analytically and experimentally the existence of ``magic'' dressing conditions where decoherence due to electromagnetic field noise is strongly suppressed. Building upon this result,  we propose a bi-chromatic dressing configuration that reduces the global sensitivity of the atomic ground states to low-frequency noise, and enables the simultaneous protection of multiple transitions between the two ground hyperfine manifolds of atomic alkali species. Our methods produce protected transitions between any pair of hyperfine sub-levels at arbitrary (low) DC-magnetic fields.  
\end{abstract}


\maketitle

\section{\label{sec:intro2}Introduction}

The sensitivity to environmental field fluctuations of atomic transitions in quantum systems represents a major challenge for improving the accuracy and reliability of applications such as atomic clocks \cite{bloom2014optical}, low-frequency field sensing \cite{PhysRevLett.116.240801}, and quantum memories \cite{zhao2009long} and information processors \cite{Preskill1998ReliableComputers,Gottesman1999Fault-TolerantSystems,sinuco2016addressed}. The sensitivity problem arises since, typically, the transition frequency between pairs of quantum states is affected by temporal and spatial variations of the electromagnetic environment, which results in a rapid dephasing of the system's wavefunction. The ubiquity of this problem, identified in ultracold atomic ensembles \cite{PhysRevLett.82.4619}, superconducting devices \cite{PhysRevB.71.094519},  nitrogen-vacancy centres in diamond \cite{PhysRevLett.112.116403} and doped silicon \cite{PhysRevB.71.014401}, has driven the development of various flavours of dynamical coherent control (pulsed \cite{PhysRevLett.90.037901}, concatenated \cite{cai2012robust} and continuous \cite{PhysRevA.85.040302,laucht2017dressed}) that  isolate such systems from unwanted noise sources, and improve their coherence time by several orders of magnitude \cite{timoney2011quantum}. In atomic systems, in particular, reduction of the frequency broadening can be achieved by applying electromagnetic fields (DC magnetic field \cite{PhysRevLett.83.3358,PhysRevA.92.012106}, microwaves \cite{Sarkany}, radio-frequency \cite{Kazakov2015}, optical radiation \cite{PhysRevLett.91.173005}) tuned to suppress the differential Zeeman or Stark energy shifts between selected pairs of states (clock/qubit states), which, effectively, protects the transition frequency against field fluctuations.   

Here we demonstrate the control of the magnetic field sensitivity of the electronic ground state of $^{87}$Rb, dressed by a radiofrequency magnetic field. The dressing is achieved by applying an oscillating magnetic field, $\boldsymbol{B}_{\textrm{RF}}$, with frequency, $\omega_{\textrm{RF}}$, close to the Larmor frequency (or Zeeman splitting) and controlled by an applied static magnetic field, $\boldsymbol{B}_{\textrm{DC}}$.  We use microwave spectroscopy \cite{PhysRevLett.24.861, sinuco2019microwave} to determine the energy difference and line-width of transitions between dressed states \cite{Garraway_2016,pandey2019hypersonic}, and study their dependence with respect to the dressing configuration. We observe that, in general, for any pair of dressed states there is a dressing condition for which the atoms decouple from fluctuations of the static field, which results in a significant reduction of the line broadening. In our setup, using a linearly polarised RF dressing field, we find that the broadening of dressed transitions lines ($\sim 0.1~$kHz) is one order of magnitude smaller than that of equivalent bare transitions ($\sim 1.0~$kHz), limited mainly by amplitude noise of our RF generator \cite{morizot2008influence}. We also find that the optimal dressing condition (i.e. that with reduced magnetic field sensitivity) depends on the selected pair of dressed states as a result of the difference of the gyromagnetic factors of the two electronic ground state hyperfine manifolds, non-linear Zeeman shifts \cite{Sinuco2012radio} and Bloch-Siegert shift effects \cite{PhysRev.57.522}. 

While the majority of existing techniques for noise suppression focuses on qubit and clock applications, there is growing interest in manipulating qudit systems in atomic \cite{JessenControl,PhysRevA.97.013407} and solid-state \cite{PhysRevLett.121.023601} platforms,  which requires the control of $d$ internal states \cite{choi2017dynamical,PhysRevLett.113.230501, lanyon2009simplifying}. In this paper, we propose a dressing scheme that reduces the sensitivity of \textit{all} possible hyperfine transitions in the electronic ground state of $^{87}$Rb. Our scheme exploits the possibility of addressing independently each hyperfine manifold by tuning the frequency of each circular component ($\sigma_{\pm}$) of the dressing field \cite{Mas2019}. 

As a figure of merit in the improvement in the stability against low-frequency noise,  and thus of the atomic resilience to decoherence \cite{PhysRevApplied.3.044009,PhysRevA.68.012311}, we evaluate the root-mean-square average of the first derivative of the atomic transitions frequencies with respect to static field, $\left\langle \alpha_{DC} \right\rangle$ (see Sec. \ref{sec:sensitivity}). Using the Rotating Wave Approximation (RWA) and neglecting the quadratic Zeeman shift, we found a first estimate of the ratio between the $\sigma_{\pm}$ frequencies that minimises $\left\langle \alpha_{DC} \right\rangle$. This optimal condition should be corrected to take into account effects from non-linear Zeeman and Bloch-Siegert shifts, which we explain qualitatively and evaluate numerically. We found that, although it is not possible to fully cancel the influence of magnetic field fluctuations for all transitions, our scheme defines dressed states with a magnetic sensitivity smaller than that possible with bare and single frequency dressed atoms. This scheme can be used to improve the robustness of qudits encoded in the electronic ground state of alkali atoms, for which control protocols have been recently demonstrated using microwave pulses  \cite{JessenControl,anderson2015accurate}.

The structure of this paper is as follows. Section \ref{sec:theory} reviews the formalism we use to define the basis of dressed states. In section \ref{sec:experiments} we present experimental results for the dependence of the transition frequencies between dressed states with respect to the applied static magnetic field, which reflects into a dependence of the linewidth and decoherence. Following this, in section \ref{sec:sensitivity} we propose a bichromatic dressing configuration that leads to an overall reduction of the magnetic  sensitivity (on average) of all possible transitions between dressed states. The closing section (\ref{sec:conclusions}) presents the conclusions of our work.


\section{\label{sec:theory} RF-dressing of the ground state manifold of alkali atoms}

The internal dynamics of an alkali atom in its electronic ground state interacting with a magnetic field,  $\boldsymbol{B}(t)$, are governed by the Hamiltonian:
\begin{equation}
H(t) = \frac{A}{\hbar^2} \boldsymbol{I}\cdot\boldsymbol{J} + \frac{\mu_{\textrm{B}}}{\hbar} (g_I \boldsymbol{I} + g_J \boldsymbol{J}) \cdot \boldsymbol{B}(t) 
\label{eq:H_generic}
\end{equation}
where $A$ is the hyperfine coupling, $\mu_{\textrm{B}}$ is the Bohr magneton and the factors $g_I$ and $g_J$ are the nuclear ($\boldsymbol{I}$) and electronic
($\boldsymbol{J}$) Land\'e g-factors, with corresponding angular momentum operators $\boldsymbol{I}$ and $\boldsymbol{J}$.
The coupling between nuclear and electronic magnetic momenta defines two hyperfine manifolds (given $J=1/2$)
with different total angular momentum, $F =I \pm 1/2$, which are split by an energy gap of $\hbar \omegaHF = A\left(I+1/2\right)$. This splitting defines a natural basis of states to describe the atomic dynamics, labelled by the total angular momentum ($F$) and its projection along a quantisation axis ($m$), $\{ \left|F,m\right\rangle \}$.  

Here we study RF-dressed atoms of $^{87}$Rb prepared by the magnetic field:
\begin{equation}
  \boldsymbol{B}(t) = B_{\textrm{DC}} \, \hat{\boldsymbol{e}}_z  +  \sum_{i\in\{x,y,z\}} B_{\textrm{RF},i} \cos(\omega_{\textrm{RF}} t + \phi_i) \hat{\boldsymbol{e}}_i
\label{eq:TotalField}
\end{equation}
where the quantisation axis $z$ is defined along the direction of the static field $\boldsymbol{B}_{\textrm{DC}}$. 

The dressed basis is defined as the set of solutions of the Schr\"odinger equation resulting from Eqs. (\ref{eq:H_generic}) and (\ref{eq:TotalField}), with the form:
\begin{equation}
\left|\Psi_{F \bar{m}} \right\rangle = e^{-i\bar{E}_{F,\bar{m}}  t/\hbar} \left| F,\bar{m} \right\rangle
\label{eq:FloquetState}
\end{equation}
where $\bar{E}_{F,\bar{m}}$ is the corresponding dressed energy \cite{shirley1965solution}. The ket on the right hand side of Eq. (\ref{eq:FloquetState}) is a time-periodic linear combination of the bare states, which can be expressed as the Fourier series:
\begin{equation}
\left|F,\bar{m}\right\rangle = \sum_{m} \sum_n U_{m,\bar{m}}^n e^{i n \omega_{\textrm{RF}}t} \left|F, m \right\rangle
\label{eq:DressedStateFourier}
\end{equation}
The existence of this solution of the Schr\"odinger equation is guaranteed by the Floquet-Bloch theorem \cite{WEINBERG20171}. 

In our experiments, see Sec. \ref{sec:experiments}, we transform the eigenstates of the static Hamiltonian, $\{\left| F,m \right\rangle\}$ into the dressed states $\{\left| F,\bar{m} \right\rangle\}$, by adiabatically varying the DC and RF fields until reaching a final dressing configuration. We achieve high fidelity transformations from the bare to the dressed basis by defining temporal trajectories of the field configuration that avoid the degeneracy of the dressed energies $\{ \bar{E}_{F,\bar{m}} \}$ \cite{WEINBERG20171}. Furthermore, the absence of level crossings during the switch-over to the dressed basis allows us to establish a one-to-one correspondence between the bare and the dressed basis, such that we can use the same set of quantum numbers for labelling the two bases.

In the limit of weak dressing and linear Zeeman shift, the Fourier coefficients in Eq. (\ref{eq:DressedStateFourier}) can be approximated using the Rotating Wave Approximation (RWA) with the dressed energy \cite{SERIES19781} (see appendix \ref{ap:appendixA}):
\begin{equation}
  \bar{E}_{F,\bar{m}}= E_F + \frac{g_F}{|g_F|} \bar{m} \sqrt{(\hbar \omega_0 - \hbar \omega_{\textrm{RF}}^{\textrm{sgn}(g_F)})^2 +  2\left|\hbar \Omega_{\textrm{RF}}^{\textrm{sgn}(g_F)}\right|^2}
\label{eq:dressedEnergyRWA}
\end{equation}
where $E_F = A(F(F+1) - I(I+1) - J(J+1))/2$, $\omega_0=|\mu_Bg_F B_{\textrm{DC}}/\hbar|$ the Larmor angular frequency, $\omega_{\textrm{RF}}^{\textrm{sgn}(g_F)}$ denotes the rotating frequency of the $\sigma_{\textrm{sgn}(g_F)}$-polar component of the RF field, defined with respect to the local direction of the static magnetic field ($z-$axis), and the Rabi coupling:
\begin{equation}
\Omega_{\textrm{RF}}^{\textrm{sgn}(g_F)} = \frac{\mu_B g_F}{2^{3/2}\hbar}\left(-\textrm{sgn}(g_F) B_{\textrm{RF},x}e^{-i\phi_x} + i B_{\textrm{RF,y}} e^{-i \phi_y}\right)
\label{eq:polarcomponents}
\end{equation} 
with $\textrm{sgn}(g_F)\in \{+1,-1\}$.

 The general definition of the dressed states, $\left|F,\bar{m}\right\rangle$ as the Fourier series in Eq. (\ref{eq:DressedStateFourier}) allows us to calculate corrections to the RWA and include the full dependence of the Zeeman shifts with the static magnetic field. This formulation can be extended to the case of polychromatic dressing with $N$ inconmensurable frequencies, using the multidimensional Fourier decomposition of the dressed state:
\begin{equation}
\left\langle F, m \right| \left. F, \bar{m}\right\rangle = \sum_{\vec{n}} U_{m,\bar{m}}^{\vec{n}} e^{i\vec{n} \cdot \vec{\omega}t} 
\label{eq:multimodeFloquetTrans}
\end{equation}
where the $N$-dimensional vectors $\vec{\omega}=(\omega_1,\omega_2, ... \omega_N)$ and $\vec{n} \in Z^N$ \cite{Sinuco2020}. We apply this formalism in Sec. \ref{sec:sensitivity} to evaluate the dressed energies when applying a bichromatic RF field. 

\subsection{Resonant condition in RF-dressed states} 

In the experiments described in Sec. \ref{sec:experiments}, we perform microwave spectroscopy of the dressed atoms by applying a short and weak MW pulse to the dressed ensemble, followed by the detection of the fraction of atoms remaining in (and transferred to) the initial (the final) hyperfine manifold \cite{sinuco2019microwave}.  

To calculate the response of the dressed atom to this pulse, we should express the MW coupling in the dressed basis, using the transformation rule: 
\begin{equation}
 \bar{H}_{\textrm{MW}} = U_F^{\dagger}(t) H_{\textrm{MW}} U_F(t) 
\end{equation} 
where $H_{\textrm{MW}}$ is the atomic coupling to the microwave field expressed in the basis of Zeeman states (see Eq. (\ref{eq:H_generic})), and $U_F(t)$ is the unitary time-dependent transformation defined in Eq. (\ref{eq:DressedStateFourier}).

The harmonic components of $U_F(t)$ combine with the oscillation of the MW field, to produce a resonant condition for transitions between dressed states, $\left|F,\bar{m}\right\rangle \leftrightarrow \left|F',\bar{m}'\right\rangle$:
\begin{equation}
\omega_{\textrm{MW}} =  n \omega_{\textrm{RF}}  + \omega_{\bar{m}', \bar{m}}  \label{eq:resonantcondition} 
\end{equation} 
with $n \in Z$ and the transition angular frequency:
\begin{equation}
\omega_{\bar{m}', \bar{m}} = \frac{\bar{E}_{F',\bar{m}'} - \bar{E}_{F,\bar{m}}}{\hbar} \label{eq:resonantcondition2} 
\end{equation} 
where the dressed energy $\bar{E}_{F,\bar{m}}$ is defined in Eq. (\ref{eq:FloquetState}).

Using the Rotating Wave Approximation and considering near-resonant RF dressing, this resonant condition becomes (see Appendix \ref{ap:appendixA}):
\begin{widetext}
\begin{eqnarray}
\omega_{\textrm{MW}}  &=&  n \omega_{\textrm{RF}} + \omega_{\textrm{hfs}} + \frac{g_{F+1}}{|g_{F+1}|}\bar{m}' \sqrt{2}|\Omega_{\textrm{RF}}^+|  - \frac{g_{F}}{|g_{F}|} \bar{m} \sqrt{2}|\Omega_{\textrm{RF}}^-| + \left(\frac{g_{F+1}}{|g_{F+1}|}\frac{\bar{m}'}{|\Omega_{\textrm{RF}}^+|} - \frac{g_{F}}{|g_{F}|}\frac{\bar{m}}{|\Omega_{\textrm{RF}}^-|}\right)\frac{\omega_\textrm{RF}^2}{2^{3/2}} \nonumber \\  &&- \left(\frac{\bar{m}' g_{F+1}}{|\Omega_{\textrm{RF}}^+|} - \frac{\bar{m}g_{F}}{|\Omega_{\textrm{RF}}^-|}\right) \omega_\textrm{RF}\frac{\mu_{\textrm{B}} B_{\textrm{DC}}}{\sqrt{2}\hbar} + \frac{1}{2^{3/2}}\left[ \frac{g_{F+1}}{|g_{F+1}|}\frac{\bar{m}' g_{F+1}^2}{|\Omega_{\textrm{RF}}^+|} -  \frac{g_{F}}{|g_{F}|}\frac{\bar{m} g_{F}^2}{|\Omega_{\textrm{RF}}^-|} \right]\left(\frac{\mu_{\textrm{B}} B_{\textrm{DC}}}{\hbar}\right)^2  \nonumber \\
\label{eq:transitionfrequencies}
\end{eqnarray}
\end{widetext}
where $\omega_{\textrm{hfs}} = (I+1/2)A/\hbar$ is the hyperfine splitting.

Experimentally, we scan the MW frequency and determine the resonant condition as a function of the static magnetic field $B_{\textrm{DC}}$. Thus, we determine the energy difference between dressed states $\hbar \omega_{\bar{m}',\bar{m}} =  (\bar{E}_{F',\bar{m}'} - \bar{E}_{F,\bar{m}})$, observing a quadratic dependence of which we give details below.


\section{\label{sec:experiments} Spectroscopy and protection of RF-dressed $^{87}$Rb}

In this section, we describe the experimental procedure to achieve decoherence-free pairs of dressed Zeeman sub-levels with an ultra-cold cloud of $^{87}$Rb atoms. We focus on the dependence of the microwave resonances with the dressing configuration, which is here controlled by the applied static magnetic field. The main components of our experimental setup are a crossed red-detuned optical trap, a pair of coils that define the quantization axis, and three pairs of coils to produce static and RF magnetic fields. For a detailed description we refer to Ref.\ \cite{sinuco2019microwave}.

We use trapped ultra-cold atoms so as not to be limited by the free-fall time of the cloud. Instead, our background collisional lifetimes close to two minutes impose no limit on detecting the target coherence times in this work, e.g. $\tau_{coherence} \approx 100$\,ms. Furthermore, using an optical dipole trap, instead of a magnetic trap \cite{Mas2019}, we can generate a state-independent trapping potential and ensure that the applied RF and DC magnetic fields are nearly homogeneous over the atomic sample. All these features of our setup enable a clean interrogation of the energy difference between RF-dressed states, limited mainly by fluctuations of the dressing field amplitude and polarization.

In our system, the quantisation axis is defined by a homogeneous DC field driven by a pair of Helmholtz coils. However, since we do not implement any magnetic compensation or shielding, both the Earth's magnetic field and fields generated by the equipment nearby,  contribute to the total static field. We determine this offset field, $\boldsymbol{B}_{\textrm{DC}}^{\textrm{offset}}$ by measuring the resonant frequency of transitions between bare states with a known applied DC field, and fitting the Breit-Rabi formula. These measurements are done in between the experimental runs that register the dressed spectrum (for details see \cite{HM2019}), and we find an offset field with  $B_{\textrm{DC},z}^{\textrm{offset}} = (-0.252 \pm 0.014)~$G and a component in the x-y plane of magnitude $B_{\textrm{DC},\perp}^{\textrm{offset}} = (0.264 \pm 0.012)~$G. This field adds to the one produced by the Helmholtz coils to define total magnetic field $\boldsymbol{B}_{DC} =  \boldsymbol{B}_{DC}^{H} + \boldsymbol{B}_{DC}^{\textrm{offset}}$. We denote its magnitude by $B_{DC}$.

\subsection{\label{sec:experiments2} Preparation of the atomic cloud}

\begin{figure}[!h]
\centering
\includegraphics[width=0.98\columnwidth]{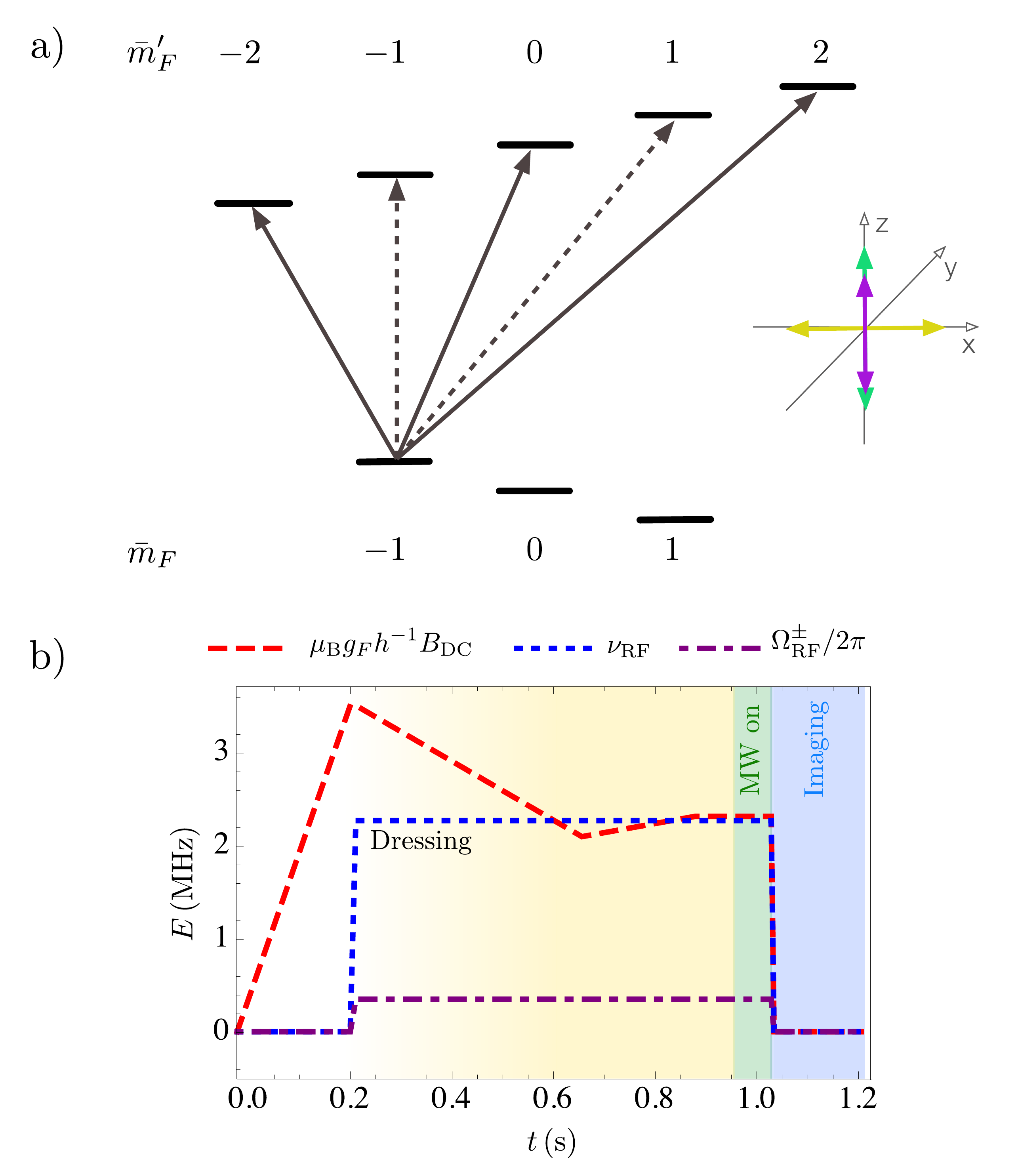}
\caption{\label{fig:dressingRamp}
(a) Schematic of the energy level diagram that indicates available microwave transitions from the initial dressed state $|1,\bar{m}_F=-1\rangle$ state to the upper $F=2$ manifold dressed sub-levels. The solid (dashed) arrow lines indicate allowed (non-allowed) transitions for the field configuration shown in the inset. Both the MW and the DC field point in the z direction (purple and green lines, respectively). The RF-field is linearly polarised along the x direction.
(b) Schematic of the experimental sequence. The red dotted line tracks the DC-field, the dashed blue line tracks the RF frequency and the purple dashed line tracks the RF-amplitude, represented by the Rabi frequency. All curves are shown in frequency units. The greyed-out areas represent the MW pulse (green) and imaging (blue). The shaded yellow are represents the adiabatic dressing of the atoms. The ramp times of $B_{\textrm{DC}},\nu_{\textrm{RF}}=\omega_{\textrm{RF}}/(2\pi)$ and $\Omega_{\textrm{RF}}$ are those described in Sec.~\ref{sec:experiments}. The extent of the ``MW on'' and ``Imaging'' greyed-out areas does not represent a real duration.}
\label{fig:experimentsequence}
\end{figure}

 Our initial sample is a cold cloud of $^{87}$Rb prepared in the bare electronic ground state $|F=1,m=-1\rangle$ (in Fig. \ref{fig:dressingRamp}(a)) via the methods described in detail in Ref.  \cite{sinuco2019microwave}. This cloud is evaporated in a hybrid crossed dipole trap plus a quadrupole (with the gradient $\alpha=10~$G/cm) until reaching a temperature of $100$\,nK with a typical population of $n=$ $2\times 10^{5}$ atoms in the bare state $|1,-1\rangle$. A bias field in the vertical direction $B_{\textrm{DC}}^{H}\approx1.2$\,G, produced by a pair of Helmholtz coils, is present at this point too. The dipole trap has been tuned at $\alpha=2~$G/cm so that the trapping frequencies are $\omega_{\rho}/2\pi\approx 180$\,Hz and $\omega_{\textrm{axial}}/2\pi\approx 30$\,Hz. We subsequently switch off the current to the quadrupole coils, leaving a residual  magnetic quadrupole field with a maximal measured field-gradient of $\alpha<0.05~$G/cm. 

In the next step, the atoms are dressed through the procedure sketched in Fig. \ref{fig:dressingRamp}(b): the current in the pair of Helmholtz coils is ramped up to increase the applied static field from $\boldsymbol{B}_{\textrm{DC}}^{H} = 0\hat{\boldsymbol{z}}~$G to $\boldsymbol{B}_{\textrm{DC}}^{H} = 5\hat{\boldsymbol{z}}~$G in $\Delta t=200~$ms. Then a linearly polarised RF-field in the $x$ direction with angular frequency $\omega_{\textrm{RF}}/2 \pi \approx 2.3~$MHz and Rabi coupling $\sqrt{2}|\Omega_{\textrm{RF}}^{\pm}|/2 \pi \approx 350~$kHz is switched on, followed by an adiabatic linear ramp  (of duration $\Delta t=0.4~$s) from $\boldsymbol{B}_{\textrm{DC}}^{H}=5\hat{\boldsymbol{z}}~$G to $\boldsymbol{B}_{\textrm{DC}}^{H}=3\hat{\boldsymbol{z}}~$G. Then, the magnitude of the Helmholtz field, $B_{\textrm{DC}}^{H}$, is increased to a final value so that the total field $B_{\textrm{DC}}$ is within the range $[3.1,3.3]~$G. Finally we probe the RF-dressed atom with a MW pulse and take quasi-simultaneous independent absorption images of the $F=1$ and $F=2$ manifolds, shortly after switching off the dipole trap that holds the cloud \cite{HM2019}. We will refer to the number of atoms measured in $F=2$ ($F=1$) as $n_2$ ($n_1$), and the fraction of atoms in $F=2$ as $f_2=n_2/(n_1+n_2)$.

\subsection{\label{sec:Resonances} Protected transitions with ultra-cold RF-dressed atoms}

\begin{figure}[!htb]
\centering
\includegraphics[width=0.48\textwidth]{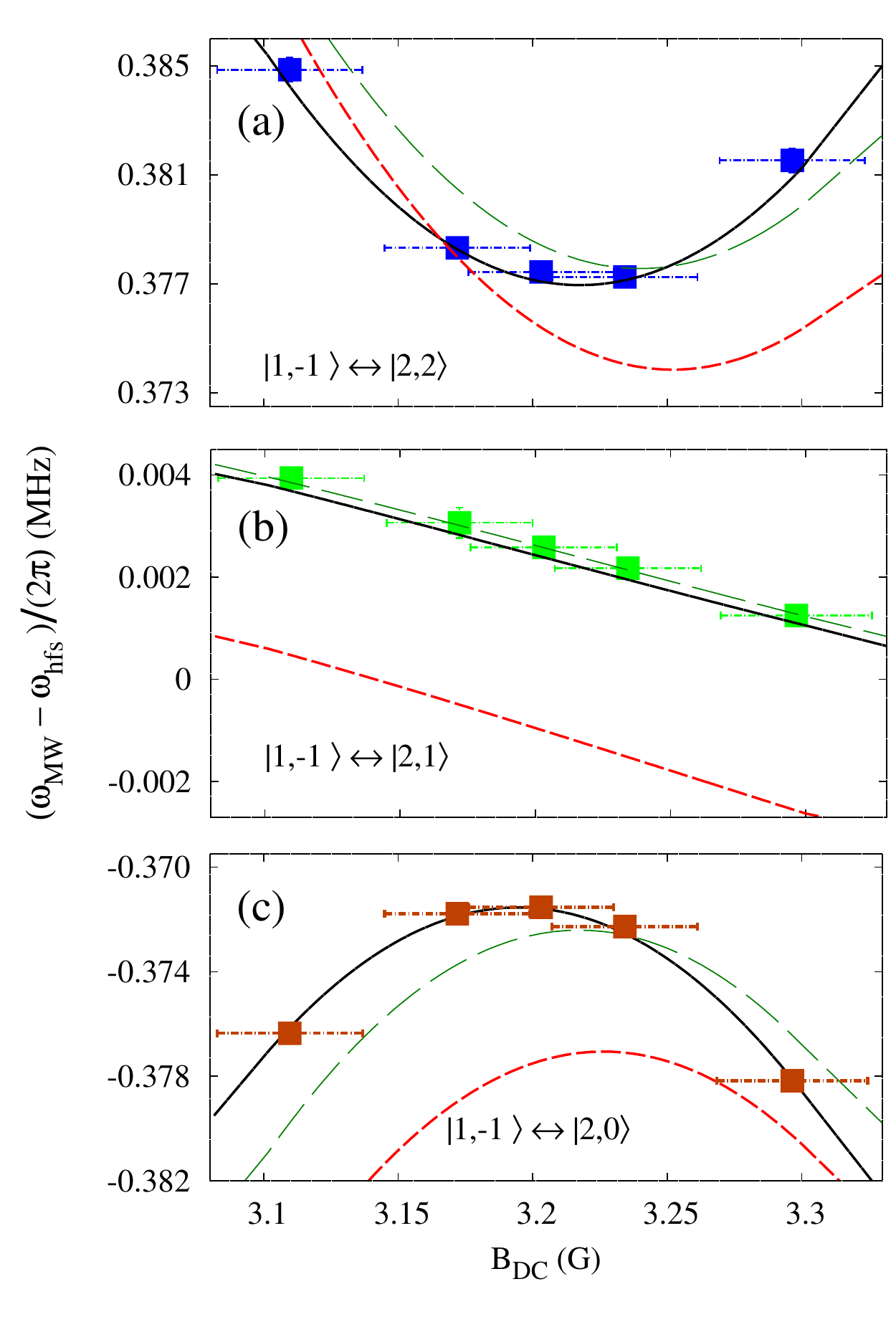}
\caption{\label{fig:theParabolas} Detuning of the resonant frequencies of the transitions between the dressed states (a) $ \left| 1,-1 \right\rangle \rightarrow \left| 2,2\right\rangle$ (b)  $\left|1,-1\right\rangle \rightarrow \left|2,-1\right\rangle$ (c) $\left|1,-1\right\rangle \rightarrow \left|2,0\right\rangle$, as functions of the applied static magnetic field. The symbols correspond to experimental values obtained from spectral signals. The solid lines are fits to our model, in Sec. \ref{sec:theory}, which takes into account non-linear Zeeman shifts and beyond RWA effects. The long-dashed (short-dashed) lines are fits to a model considering the RWA and non-linear (linear) Zeeman shifts. The amplitude of the RF field  is the only free parameter of our models since we measure the total static magnetic field. For the full model (long-dashed) we obtained $B_{\textrm{RF}}^{+} =  (1.07 \pm 0.09)~$G, while for for the solid lined $B_{\textrm{RF}}^{+} = (1.08 \pm 0.09)~$G, with $B_{\textrm{RF}}^{\textrm{sgn}(g_F)} = \sqrt{2}|\Omega_{\textrm{RF}}^{\textrm{sgm}(g_F)}/(\mu_B g_F)|$ The short-dashed line is evaluated using $B_{\textrm{RF}}^+ =   1.075~$G and the measured DC magnetic field.}
\end{figure}

After preparing the atomic cloud in the dressed state $|1,\bar{m}=-1\rangle$, we fix the dressing frequency $\omega_{\textrm{RF}}^{\pm}/2 \pi=2.263410$\,MHz and Rabi couplings $\sqrt{2}|\Omega_{\textrm{RF}}^{\pm}|/2 \pi \approx 350~$kHz. Then we determine the resonant frequency of three transitions in the vicinity of the zero field hyperfine splitting \cite{sinuco2019microwave} as functions of the applied static field $B_{\textrm{DC}}^{H}$. For each value of the total static magnetic field, $B_{\textrm{DC}}$, we extract the resonant transition frequency after fitting Lorentzian curves to measurements of the transferred atomic state population following a short microwave pulse and scanning the microwave frequency. The transitions investigated have the final states $|2,\bar{m}=1\rangle$ (at $\omegaMW \approx \omegaHF$), $|2,\bar{m}=2\rangle$ (at $\omegaMW \approx  \omegaHF+\sqrt{2}|\Omega_{\textrm{RF}}^+|$) and $|2,0\rangle$ (at $\omegaMW \approx \omegaHF-\sqrt{2}|\Omega_{\textrm{RF}}^+|$), and the corresponding resonant frequencies can be estimated from Eq.~(\ref{eq:resonantcondition}). Results of these measurements are shown in Fig. \ref{fig:theParabolas}. When sampling the total field $B_{\textrm{DC}}$ near the resonant condition of the RF field with the Larmor frequency, we observed a linear dependence of the resonant transition frequency to the state $|2,\bar{m}=1\rangle$, and a quadratic dependence for the transitions to $|2,\bar{m}=2\rangle$ and $|2,\bar{m}=0\rangle$, which is in qualitative agreement with Eqs.~(\ref{eq:dressedEnergyRWA})-(\ref{eq:resonantcondition}) after taking into account the difference between Land\'e factors of the two hyperfine manifolds. Even though we represented the transition to the $|2,1\rangle$ state as non-allowed in Fig.\ref{fig:experimentsequence}, this occurs only in the idealised case where all fields are aligned as the inset of this figure indicates, see Ref. \cite{sinuco2019microwave}. 

We also observe that the transition lines from the dressed state $|1,\bar{m}=-1\rangle$ to $|2,\bar{m}=0\rangle$ and $|2,\bar{m}=2\rangle$  narrow down to a linewidth of only $\Delta \nu \approx 110~$Hz (for a MW pulse duration of $\Delta t=10~$ms) when the static field is adjusted to the extrema of the quadratic line-shifts, even without using active stabilization of  $B_{\textrm{DC}}$ or magnetic shielding. 

\begin{figure}[!htb]
\centering
\includegraphics[width=0.4\textwidth]{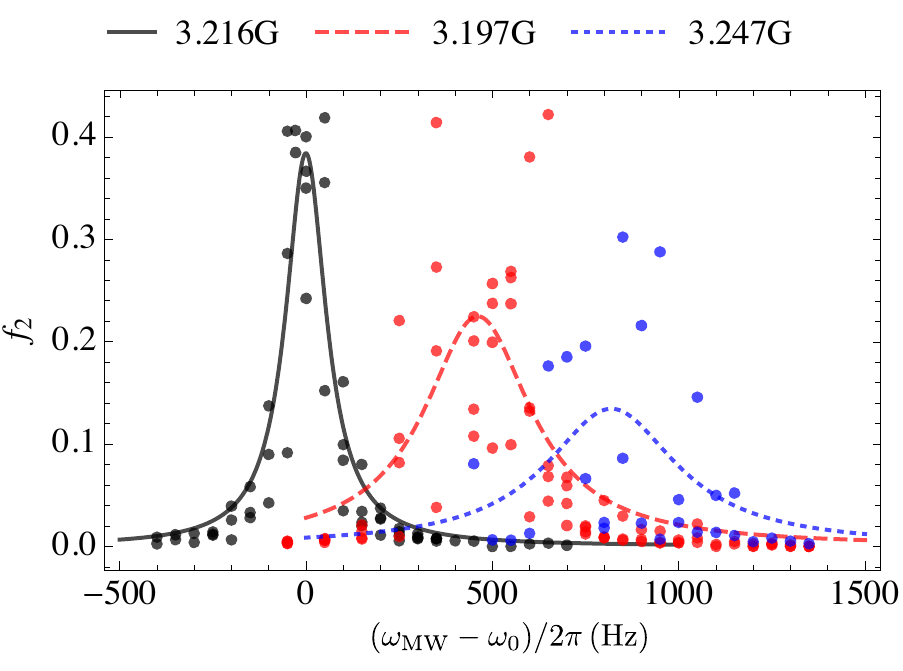}
\caption{\label{fig:thenarrowing}
Experimental demonstration of line narrowing in the dressed RF system. 
The measurements (dots) of the transition $|1,\bar{m}=-1\rangle \rightarrow |2,\bar{m}=2\rangle$ for three different total fields $B_{\textrm{DC}}$: $3.195~$G (red), $3.247~$G (blue) and $3.216~$G (black). This last value corresponds to the minimum of the parabola in Fig.~\ref{fig:theParabolas}(a). In all cases the MW pulse duration is $\Delta t=10$\,ms. The horizontal axis is the detuning $\Delta \omega_{0}=\omega_{\textrm{MW}}-\omega_{0}$ with respect to the minimum of the parabola, which we have labeled $\omega_{0}$. The vertical axis shows the fraction of measured population in $F=2$, that we defined as $f_{2}$ in \ref{sec:experiments2}. The lines are fits to Lorentzian curves for the three total magnetic fields: $B_{\textrm{DC}}=3.195~$G (dashed, red), $B_{\textrm{DC}}=3.247~$G (dotted, blue) and $B_{\textrm{DC}}=3.216~$G (solid, black). One can see that the randomness of the transition increases as one interrogates further away from the condition tha minimize the parabola in Fig.~\ref{fig:theParabolas}(a). This is due to the higher sensitivity to DC magnetic field fluctuations. The fitted Lorentzian linewidths are: $\Delta \nu = (132\pm 13)~$Hz, $\Delta \nu = (347\pm 54)~$Hz and $\Delta \nu = (427\pm 189)~$Hz for the black, red, and blue curves, respectively.}
\end{figure}

In Fig.~\ref{fig:thenarrowing} we show the line-shape of the transition $|1,\bar{m}=-1\rangle\rightarrow |2,\bar{m}=2\rangle$ for three particular total fields $B_{\textrm{DC}} = 3.197~$G, $3.216~$G and $3.247~$G. This figure shows the fraction of atoms transferred to the $F=2$ manifold as a function of the applied microwave frequency, in the vicinity of the hyperfine splitting frequency, $\omega_{\textrm{hfs}}$. The most striking feature of these measurements is the narrowing of the line-shape as the static field approaches the turning point of the quadratic line-shift in Fig. \ref{fig:theParabolas}(a), where the transition becomes protected against fluctuations of the DC field.

\begin{figure}[!htb]
\centering
\includegraphics[width=0.43\textwidth]{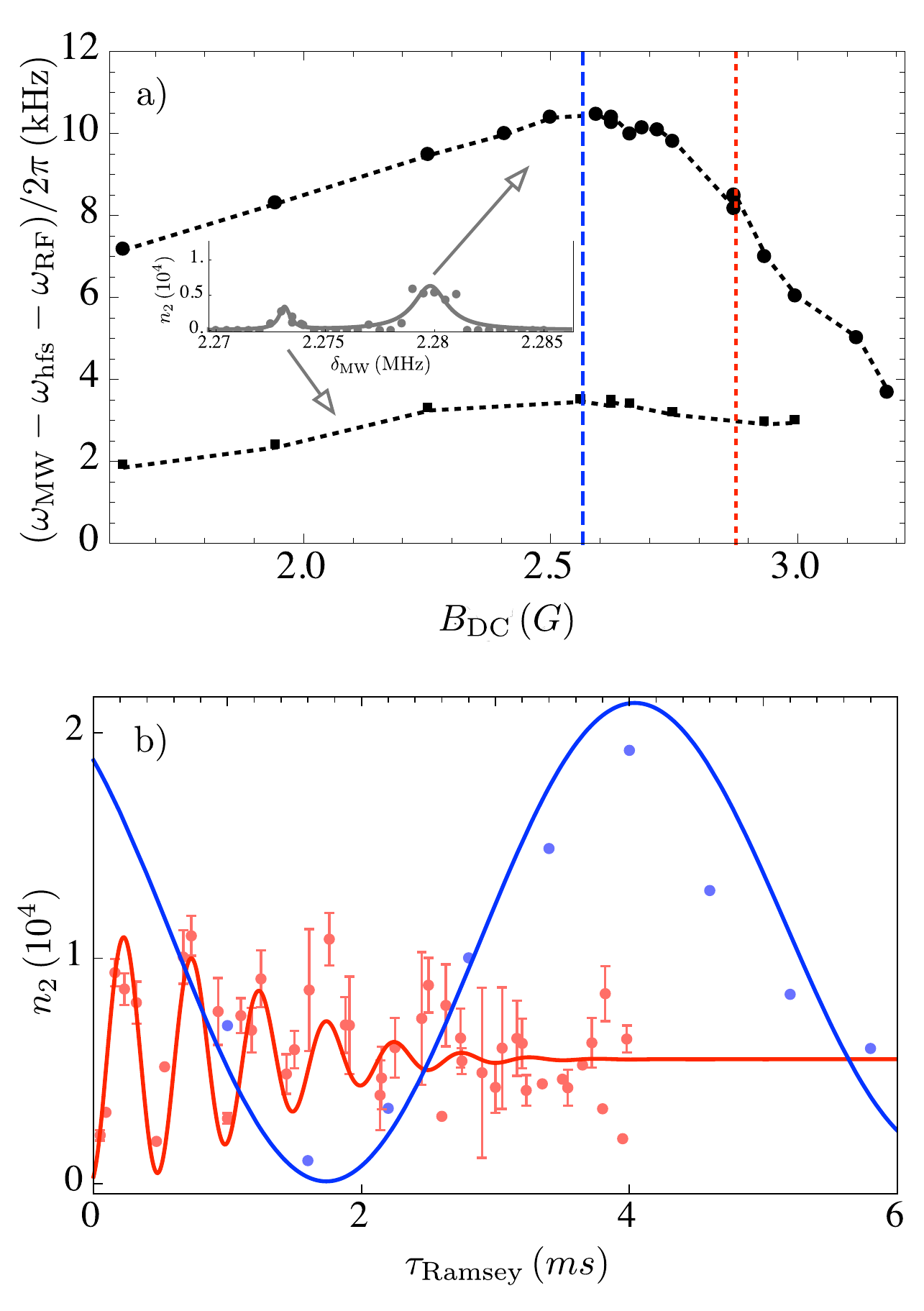}
\caption{\label{fig:ramsey2dcfields} Line-shift and Ramsey fringes observed in $^{87}$Rb dressed with the Rabi frequency  $\Omega_{\textrm{RF}}/ 2 \pi = 300~$KHz and the RF frequency  $\omega_{\textrm{RF}}/ 2 \pi = 2.27~$MHz, as functions of the static magnetic field.  (a) Shows the line-shift for the $\left|1,\bar{m}=-1\right\rangle \rightarrow \left|2,\bar{m}=1\right\rangle$ transition. We observe a second weaker peak in the range of the static field explored (see Appendix~\ref{ap:appendixC}). Inset: shows a typical spectral signal with two Lorentzian peaks matched to data in the main panel. (b)  Shows Ramsey fringes of the atomic population of the upper hyperfine manifold, $n_{F=2}$, when $B_{\textrm{DC}}=2.87~$G (red line) and $B_{\textrm{DC}}= 2.56~$G (blue line). The vertical dashed and dotted lines in (a) indicate the values of the magnetic fields used to obtain the Ramsey fringes in (b). Also in (b), the red dots with errors bars are RMS values and standard deviations, respectively, obtained from 5 repetitions per point. Red and blue dots without error bars are single measurements of the total number of atoms in the $F=2$ manifold, $n_2$.  To highlight the contrast between the noise-sensitive (red dots) and the protected (blue data) dressing configurations,  we only show the first $65~$ms of a longer experimental run (not shown).  Considering Ramsey fringes modelled with exponential decay, the protected configuration (blue dots) displays a dephasing time of $\tau=17~$ms while the noise-sensitive case shows a dephasing time of  $\tau=1.65~$ms.}
\end{figure}

Our data in Fig. \ref{fig:theParabolas}(b) shows that the transition to $|2,\bar{m}=1\rangle$ presents a remarkably different behaviour compared to the other two. In this case, the resonant frequency presents a linear dependence with the static field in the vicinity of the single MW photon transition (i.e. with $n=0$ in Eq. (\ref{eq:resonantcondition})),  preventing us from finding a condition to null the DC field sensitivity of this transition. However, as shown in Fig. \ref{fig:ramsey2dcfields}(a), the resonant frequency displays a turning point when the magnitude of the DC field enables a two-photon transition, which corresponds to the resonant condition with $n=1$ in Eq. (\ref{eq:resonantcondition}) \cite{sinuco2019microwave}. Under this condition, we consistently measured coherence times $>50$\,ms in a Ramsey type experiment in the crossed-dipole trap, which quickly shortens when setting the DC field away from the turning point \cite{merkel2019magnetic}.

Figures \ref{fig:ramsey2dcfields}(a)-(b) illustrate the behaviour of the transition $\left|1,\bar{m}=-1\right\rangle \rightarrow \left|2,\bar{m}=1\right\rangle$ far from the resonant point of the RF frequency with the Larmor frequency and with atoms trapped in a crossed-dipole potential, as before. The panel (a) of Fig.~\ref{fig:ramsey2dcfields} shows the dependence with the static field of the detuning of the resonant frequency defined as $\delta_{\textrm{MW}}=(\omega_{\textrm{MW}}-\omegaHF-\omegaRF)/2\pi$, where $\omegaHF$ is the hyperfine splitting frequency and $\omega_{\textrm{RF}}/2\pi=2.27~$MHz. The black dots correspond to the center of the Lorentzian fits in Fig. \ref{fig:ramsey2dcfields}, and the black dotted line is included as a guide to the eye. We also observe the intermittent appearance of a second, weaker peak (see Appendix~\ref{ap:appendixC}) marked by grey squares and a dashed line. This feature comes from transitions from the bare state $\left|1,1 \right\rangle$, which becomes populated by non-adiabatic effects when sweeping the applied DC field \cite{PhysRevA.96.023429}. In the same panel, the vertical dashed line (blue) indicates the turning point of the curve, where $\partial \delta_{\textrm{MW}}/\partial B_{\textrm{BC}} = 0$, and the vertical dotted line (red) points to a static field with a comparatively large gradient of $\delta_{\textrm{MW}}$.

Fig.~\ref{fig:ramsey2dcfields}(b) shows two sets of data in a Ramsey-type experiment. The blue set of data is taken at the turning point of the inverted parabola in panel (a), corresponding to  $B_{\textrm{DC}}=2.56~$G. The red set of data is taken with $B_{\textrm{DC}}=2.87~$G, i.e. where the vertical red dotted line indicates in (a). For these experiments, we detune the driving MW pulse from each transition by $\delta \nu_{R}=\nu_{\textrm{MW}}-\nu_{0}$ -where $\nu_0$ is the transition frequency, and $\nu_{\textrm{MW}}$ is the MW frequency- and then drive the transition off-resonantly with two MW pulses separated by $\tau_{\text{Ramsey}}$. We then scan $\tau_{\text{Ramsey}}$ and observe the free temporal evolution of the population in the upper dressed state $\left|2,1\right\rangle$. In both cases we fit the atom number to  $f_{2}=(n_{2}/2)\left[1+  e^{-\frac{t^{2} \sigma^{2}}{2}}   \cos{\left(2 \pi t \delta_{\textrm{R}}  \right) }\right]$, where $\sigma$ is the standard deviation of a Gaussian-shaped noise profile associated with the two-level system defined by the states $\left|1,\bar{m}=-1\right\rangle$ and $\left|2,\bar{m}=1\right\rangle$. The fit to the blue set of data gives $\delta_{\textrm{R}}=216\pm2$\,Hz and $\sigma=79\pm8$\,Hz, whereas the fit to the red set of data gives  $\delta_{\textrm{R}}=1972 \pm 35$\,Hz and $\sigma=883\pm112$\,Hz. The MW pulse detuning in the red (blue) set of data was chosen to be $\delta_{\textrm{R}}=2000$\,Hz ($\delta_{\textrm{R}}=200$\,Hz), which agrees with the fit estimate. The difference in dephasing rates of the two Ramsey oscillations is a result of the distinct DC-field noise sensitivity of the dressed transitions (quantified by the gradient $\partial \delta_{\textrm{MW}}/\partial B_{\textrm{BC},z}$) at the chosen static fields on Fig.~\ref{fig:ramsey2dcfields}(a). The different amplitude of the oscillations is due to the relative shift of the driving field detuning with respect to resonant driving, which in the case of the red curve it is as much as $2~$kHz.

In both cases, the fluctuations of the  RF-field still shift the transitions and contribute to the broadening of the line-shape. For instance, in the  transition $\left| 1,\bar{m}=-1 \right\rangle \rightarrow \left|2,\bar{m}=1\right\rangle$, we measured a line shift of $7~$Hz per kHz of Rabi frequency caused by the RF amplitude noise. It is evident in any case that the DC-noise fluctuations result in a much larger dispersion of the measurements as the lines are further detuned from the extreme of the parabola-shaped line-shift. This produces a better Lorentzian fit for the black data set in Fig.~\ref{fig:thenarrowing}, and worse fits for the blue and red data sets. Both Fig.~\ref{fig:thenarrowing} and Fig.~\ref{fig:ramsey2dcfields}(b) show that transitions between Zeeman sub-states can be narrowed down in the RF-dressed regime by choosing field parameters where the sensitivity to DC-fields of the differential energy shifts become minimal, although the optimal configuration depends on the pair of states involved.

\section{\label{sec:sensitivity} Simultaneous DC protection of multiple dressed transitions}

The protection against DC-field fluctuations demonstrated in the previous section (Sec. \ref{sec:experiments}) occurs at different DC-fields for different transitions. In particular, we observed how the dressing configuration can be adjusted to reduce strongly the sensitivity of the dressed transitions $\left|F=1,\bar{m}=-1\right\rangle \rightarrow \left|F=2, \bar{m}=2 \right\rangle$ (see Fig. \ref{fig:thenarrowing}) and $\left|F=1,\bar{m}=-1\right\rangle \rightarrow \left|F=2, \bar{m}=1 \right\rangle$ (see Fig. \ref{fig:ramsey2dcfields}). This configuration is possible thanks to the nearly identical dependence of the dressed energies involved with the applied static field. These dependencies result in a condition where, to first order, variations of the static field do not affect the transition angular frequency:
\begin{equation}
\left. \frac{\partial \omega_{\bar{m}',\bar{m}}}{\partial B_{\textrm{DC}}} \right|_{B_{\textrm{DC}}^0}= 0.
\label{eq:firstorder}
\end{equation}
at a particular value of the static field, $B_{\textrm{DC}}^0$. In this section, we propose to use a combination of two RF fields to manipulate the dressed energy of the hyperfine manifolds $F=1$ and $F=2$ independently of each other, and investigate how this scheme can be used to reduce the DC-magnetic field sensitivity of several transitions at the same value of the static field.

Considering only the RWA, the upper and lower hyperfine manifolds are dressed independently by different circular component of the RF dressing field. Also, assuming that a static magnetic field of amplitude $B_{\textrm{DC}}^{0}$ produces a linear Zeeman shift, we find that all transitions between dressed states become protected when the frequency of each polar component is set according to:
\begin{eqnarray}
\hbar \omega_{\textrm{RF}}^{\textrm{sgn}(g_{F+1})} &=&  \mu_{\textrm{B}} |g_{F+1}| B_{\textrm{DC}}^0 \label{eq:resonanceUpper} \\
\hbar \omega_{\textrm{RF}}^{\textrm{sgn}(g_F)} &=&  \mu_{\textrm{B}} |g_{F}| B_{\textrm{DC}}^0 \label{eq:resonanceLower} 
\end{eqnarray}
i.e., the resonant dressing of each hyperfine manifold ensures that Eq. (\ref{eq:firstorder}) is satisfied \textit{exactly} for all transitions.

Furthermore, the RWA also tells us that when the polar components of the dressing field satisfy the condition:
\begin{equation}
\frac{B^{\textrm{sgn}(g_{F+1})}_{\textrm{RF}}}{B^{\textrm{sgn}(g_F)}_{\textrm{RF}}} = \frac{\bar{m}}{\bar{m}'}  \frac{g_{F}}{g_{F+1}}, \label{eq:secondOrderRatio}
\end{equation}
with $B_{\textrm{RF}}^{\textrm{sgn}(g_F)} = \sqrt{2}|\Omega_{\textrm{RF}}^{\textrm{sgn}(\pm)}/(\mu_B g_F)|$, all transitions between states with the same sign of the magnetic moment are also stable to second order:
\begin{equation}
\left.\frac{\partial^2 \omega_{\bar{m}',\bar{m}}}{\partial B_{\textrm{DC}}^2} \right|_{B_{\textrm{DC}}^0}= 0
\label{eq:secondorder}
\end{equation}

\begin{figure}[!htb]
\centering
\includegraphics[width=0.235\textwidth]{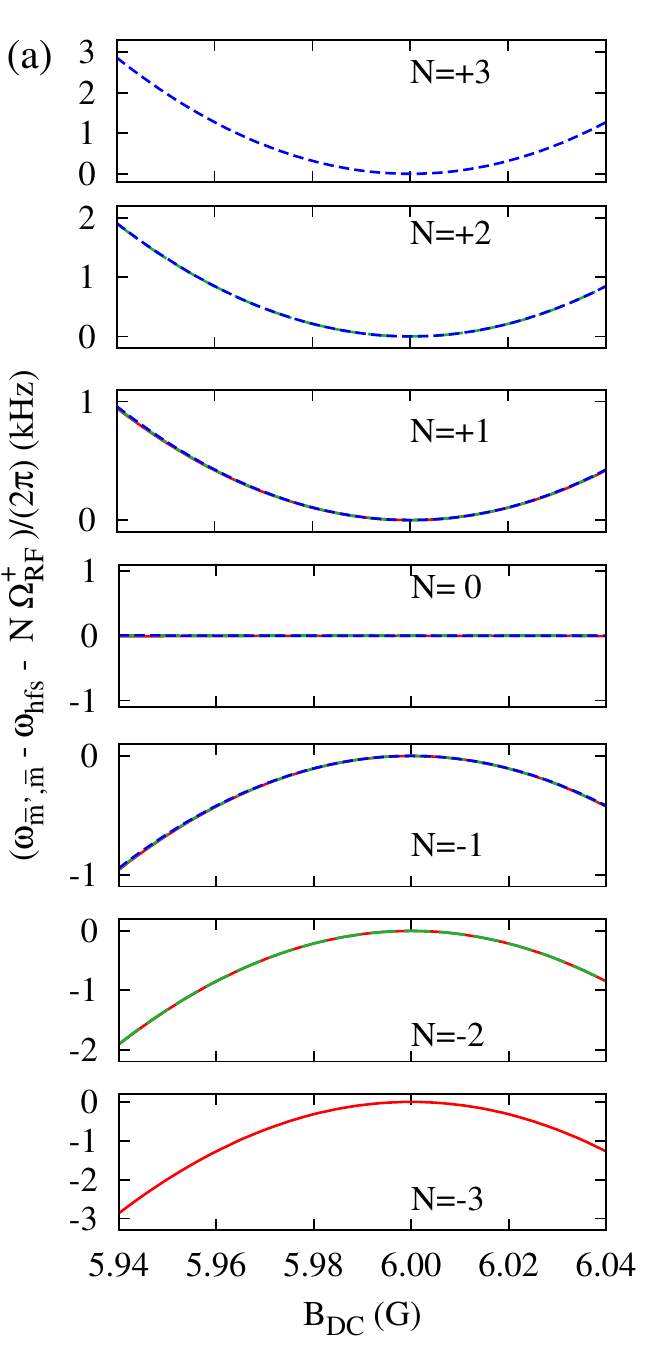}
\includegraphics[width=0.235\textwidth]{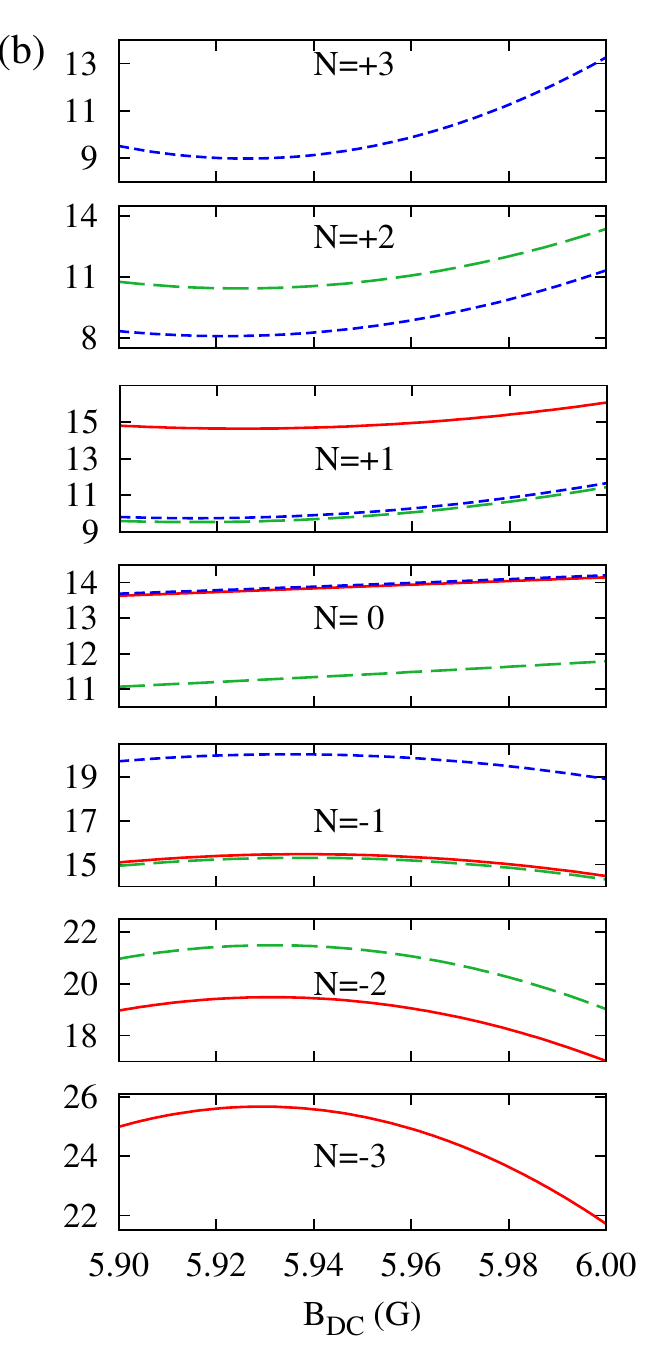}
\caption{\label{fig:threestates2}  Resonant frequencies of all hyperfine transitions in RF dressed $^{87}$Rb, as functions of the static magnetic field. An offset is applied to each resonant frequency to define a simple vertical scale with a similar range for all cases. In all panels, the dressing field configuration is given by Eqs.  (\ref{eq:resonanceUpper}) - (\ref{eq:secondOrderRatio}), with $B_{\textrm{DC}}^0=6.0~$G, $\bar{m}=-\bar{m}'=1$ and the $B_{\textrm{RF}}^+=0.12~$G. The dressed energies are calculated using (a) the RWA and (b) the  dressing scheme in Eq. (\ref{eq:multimodeFloquetTrans}) with two frequencies. Solid, dashed and short-dashed lines correspond to transitions with the initial states $|F=1,\bar{m} \rangle=$ $|1,-1\rangle$, $|1,0\rangle$, $|1,1\rangle$, respectively. The labels indicate the final states as $|F=2,\bar{m}'= N - \bar{m} \rangle$. In panel (b), the non-linear Zeeman shift and beyond RWA effects break the degeneracy of the transitions observed in panel (a). }
\end{figure}

The mechanism behind the reduction of the sensitivity of multiple transitions is shown in Fig. \ref{fig:threestates2}, where we plot the detuning of the resonant frequencies of all 15 possible transitions, $\delta_{\bar{m}',\bar{m}} = (\omega_{\bar{m}',\bar{m}} - \omega_{\textrm{hfs}})/2\pi$, as functions of the DC field. To see the effect, in Fig. \ref{fig:threestates2}(a), we use $B_{\textrm{DC}}^0=6.0~$G and $B_{\textrm{RF}}^+=0.12~$G along  with Eqs. (\ref{eq:resonanceUpper}) - (\ref{eq:secondOrderRatio}) for the frequencies and amplitudes of the circular components of the RF field. In this case, the resonant frequency of all transitions, $\delta_{\bar{m}',\bar{m}}$,  display one equilibrium point at our chosen value $B_{\textrm{DC}}^0=6.0~$G  (i.e.~$\left.\partial \delta_{\bar{m}',\bar{m}}/\partial B_{\textrm{DC}}  \right|_{B_{\textrm{DC}}^0}= 0, ~ \forall \{\bar{m}',\bar{m}\}$), which corresponds to ensuring first order (i.e. linear) protection against noise in the static field. In Fig. \ref{fig:threestates2}(b), we plot the dressed energy calculated taking into account non-linear Zeeman shifts and beyond RWA effect, using the method outlined in Sec. \ref{sec:theory}. In contrast to the previous case, now the equilibrium points of $\delta_{\bar{m}',\bar{m}}$ occur at different values of the static field, significantly distinct from $B_{\textrm{DC}}$. We conclude that, in general, the conditions Eqs. (\ref{eq:resonanceUpper})-(\ref{eq:secondOrderRatio}), do not protect the states against fluctuations of the DC field at the applied field $B_{\textrm{DC}}$ because of non-linear effects breaking the regularity of the energy spectrum assumed by our RWA model. Note, however, that with these conditions all resonant conditions experience a similar, but not identical, shift towards  $B_{\textrm{DC}}\approx 5.93~$G. 

\begin{figure}[!ht]
\centering
\includegraphics[width=0.48\textwidth]{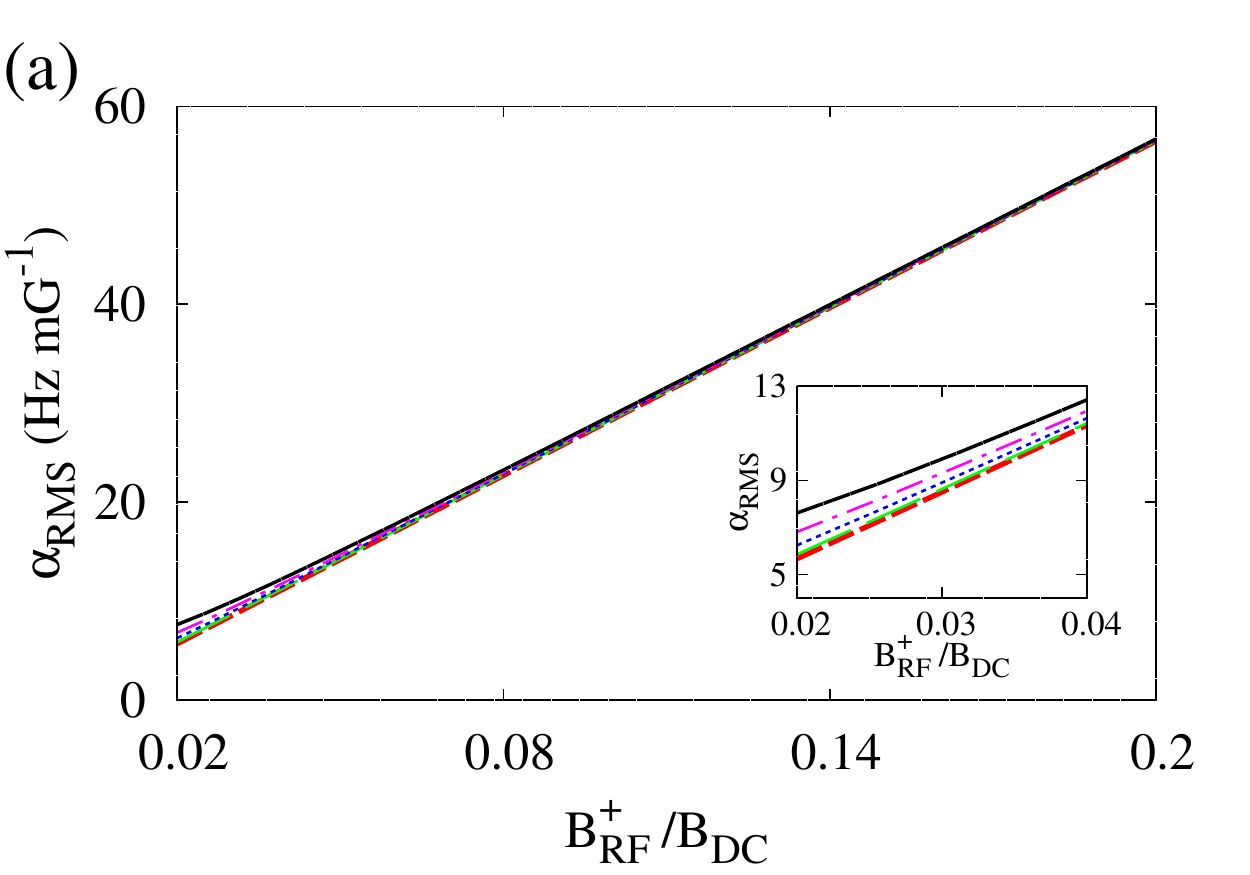}
\includegraphics[width=0.48\textwidth]{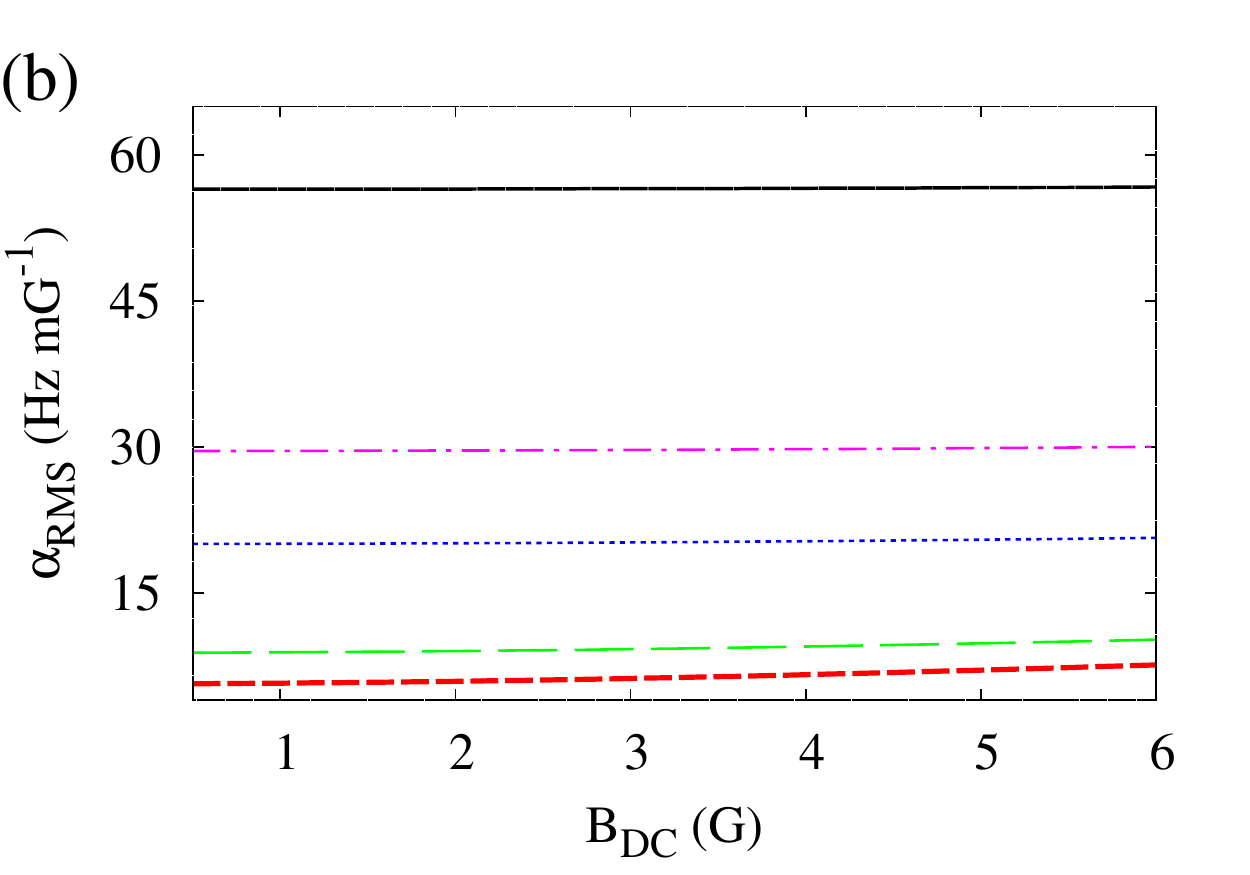}
\caption{\label{fig:RMSdifferentialmagneticmomentRWA} 
Average sensitivity $\left\langle \alpha_{\textrm{DC}} \right\rangle$ of the dressed atomic transitions with respect to the static field (Eq. (\ref{eq:alphaRMS})), calculated using  the field configuration given by the RWA Eqs. (\ref{eq:resonanceUpper}) -(\ref{eq:secondOrderRatio}). Panel (a) shows $\left\langle \alpha_{\textrm{DC}} \right\rangle$ as a function of the ratio $B_{\textrm{RF}}^{+}/B_{\textrm{DC}}$ for the DC fields $0.5~$G (short-dashed, red), $1.70~$G (dashed, green), $3.08~$G (dotted, blue), $4.28~$G (dot-dashed, cyan) and $6.0~$G (solid, black). The inset shows the expanded region of weak fields. Panel (b) shows $\left\langle \alpha_{\textrm{DC}} \right\rangle$ as a function of the static field, $B_{\textrm{DC}}$, for  RF fields of amplitude $0.5~$G (short-dashed, red), $1.70~$G (dashed, green), $3.08~$G (dotted,blue), $4.28~$G (dashed-dot, cyan) and $6.0~$G (solid, black). }
\end{figure}

We quantify the average DC sensitivity of the dressed atom using the root-mean-square (RMS) of the first derivative of the 15 microwave transition frequencies with respect to the static field:
\begin{equation}
\left\langle\alpha_{\textrm{DC}}\right\rangle  = \frac{1}{2\pi}\sqrt{ \frac{1}{15} \sum_{\bar{m}=-1}^1\sum_{\bar{m}'=-2}^2 \left(\frac{\partial \omega_{\bar{m}',\bar{m}}}{\partial B_{\textrm{DC}}} \right)^2}
\label{eq:alphaRMS}
\end{equation}
The dressed transition frequencies in Fig. \ref{fig:threestates2}(b) indicate that the non-linear Zeeman shifts and beyond RWA effects frustrate the exact first order stability expected when applying the RWA when we obtain $\left\langle \alpha_{\textrm{DC}} \right\rangle = 0$.

Setting the dressing frequencies and amplitudes according to Eqs. (\ref{eq:resonanceUpper})-(\ref{eq:resonanceLower}), we evaluated numerically $\left\langle\alpha_{\textrm{DC}}\right\rangle$ for a typical range of experimentally relevant parameters. Our numerical results, in Fig. \ref{fig:RMSdifferentialmagneticmomentRWA}, show that, when using the conditions indicated by the RWA and for sufficiently strong RF fields (e.g. $B_{\textrm{RF}}^{+} > ~ 0.08 B_{\textrm{DC}}$), the atomic sensitivity $\left\langle \alpha_{\textrm{DC}}\right\rangle$ depends linearly on the amplitude of the RF field and is independent of the applied static field. For weaker RF fields, $ \left\langle \alpha_{\textrm{DC}} \right\rangle$ depends on the static field since both beyond-RWA and non-linear Zeeman effects become comparable. 

The residual sensitivity observed in Fig.~\ref{fig:RMSdifferentialmagneticmomentRWA} can be reduced by adjusting the dressing frequencies to bring them back into resonance after taking into account energy shifts induced by the driving. Such energy shifts occur because, while each circular component of the RF field dresses one hyperfine manifold, they also cause off-resonant perturbations of the other one \cite{PhysRev.57.522}. Using a second order perturbative expansion of the RWA dressed energy, these energy shifts translate into a correction of the resonant condition given by (see Appendix \ref{ap:appendixRFNoise}):
\begin{eqnarray}
\hbar \Delta \omega_{\textrm{RF}}^{\ell_F} &=&  \frac{1}{2} \frac{|\mu_{\textrm{B}} g_{F} B^{-\ell_F}_{\textrm{RF}}|^2}{\mu_{\textrm{B}} |g_{F}|B_{\textrm{DC}} + \hbar \omega_{\textrm{RF}}^{-\ell_F}} 
\label{eq:plusCorrectionT}
\end{eqnarray}
with $\ell_F = \textrm{sgn}(g_F)$ and $F=I\pm1/2$.

\begin{figure}[!ht]
\centering
\includegraphics[width=0.48\textwidth]{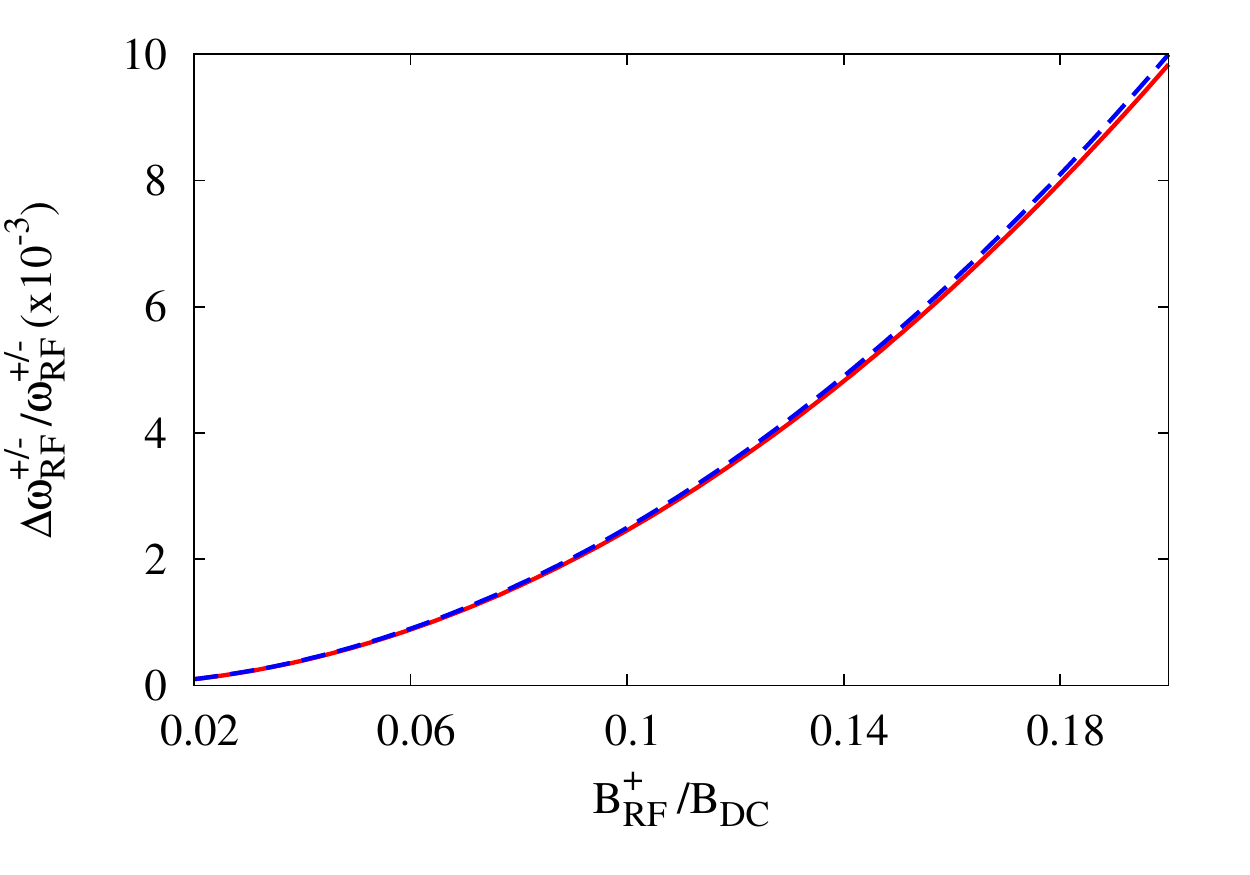}
\caption{\label{fig:FrequenciesCorrection} 
Fractional correction of the frequencies required to reduce the average atomic linear sensitivity with respect to the static field: frequency of the  $\sigma_{\textrm{sgn}(g_{F+1})}$ (dashed) and  $\sigma_{\textrm{sgn}(g_{F})}$ (solid) polar components of the dressing fields. This shift is independent of the applied static field.}
\end{figure}

With these arguments as a guide for our calculations, we numerically optimise the combination of frequencies that minimise $\left\langle\alpha_{\textrm{DC}}\right\rangle$, using field amplitudes at the  ratio given by Eq. (\ref{eq:secondOrderRatio}). The fractional frequency shift with respect to the conditions given by the RWA is shown in Fig. \ref{fig:FrequenciesCorrection}, which is in qualitative agreement with Eqs. (\ref{eq:plusCorrectionT}). Note that, although the relative frequency shift is of order $\sim 10^{-3}$, this translates into an important correction in absolute terms, typically corresponding to a shift of a few kHz of the RF frequency.

\begin{figure}[!ht]
\centering

\includegraphics[width=0.48\textwidth]{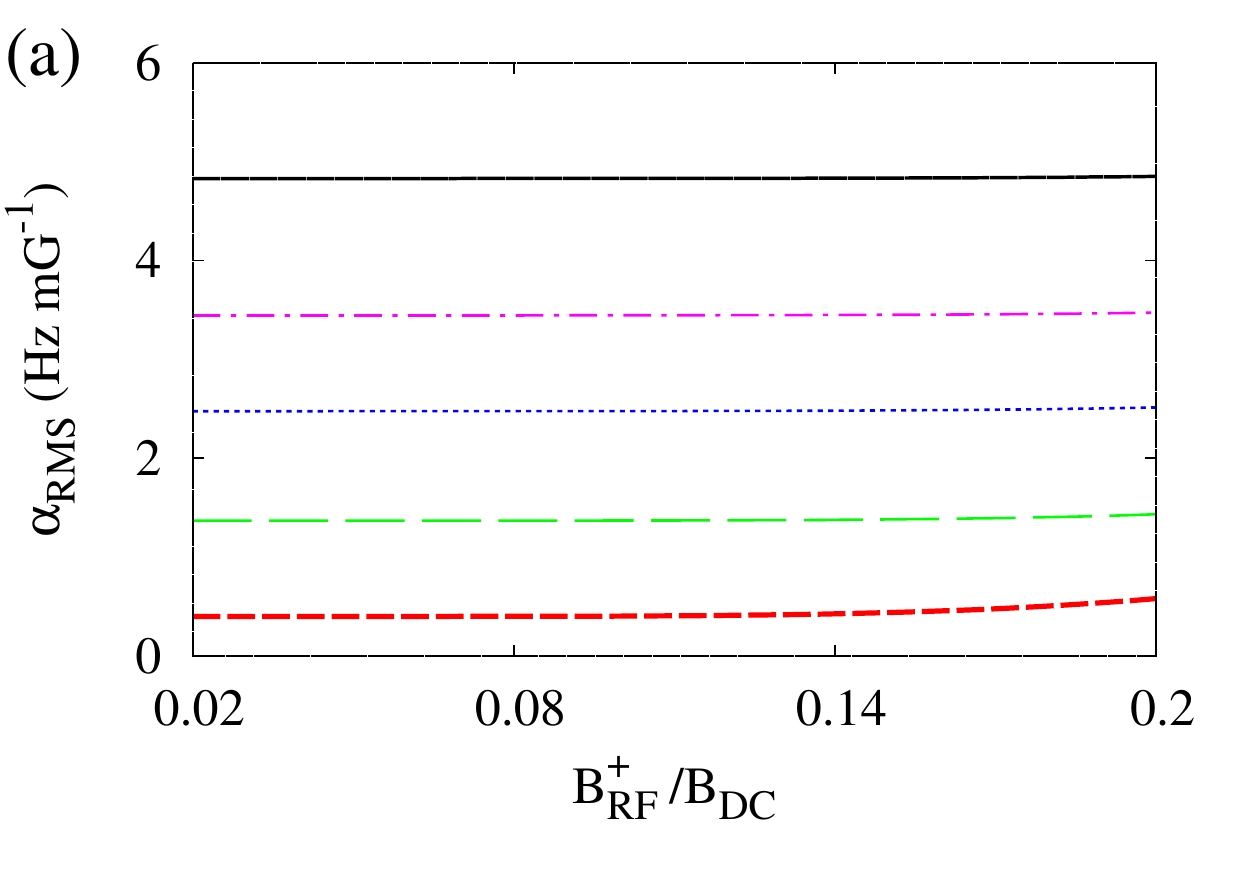}
\includegraphics[width=0.48\textwidth]{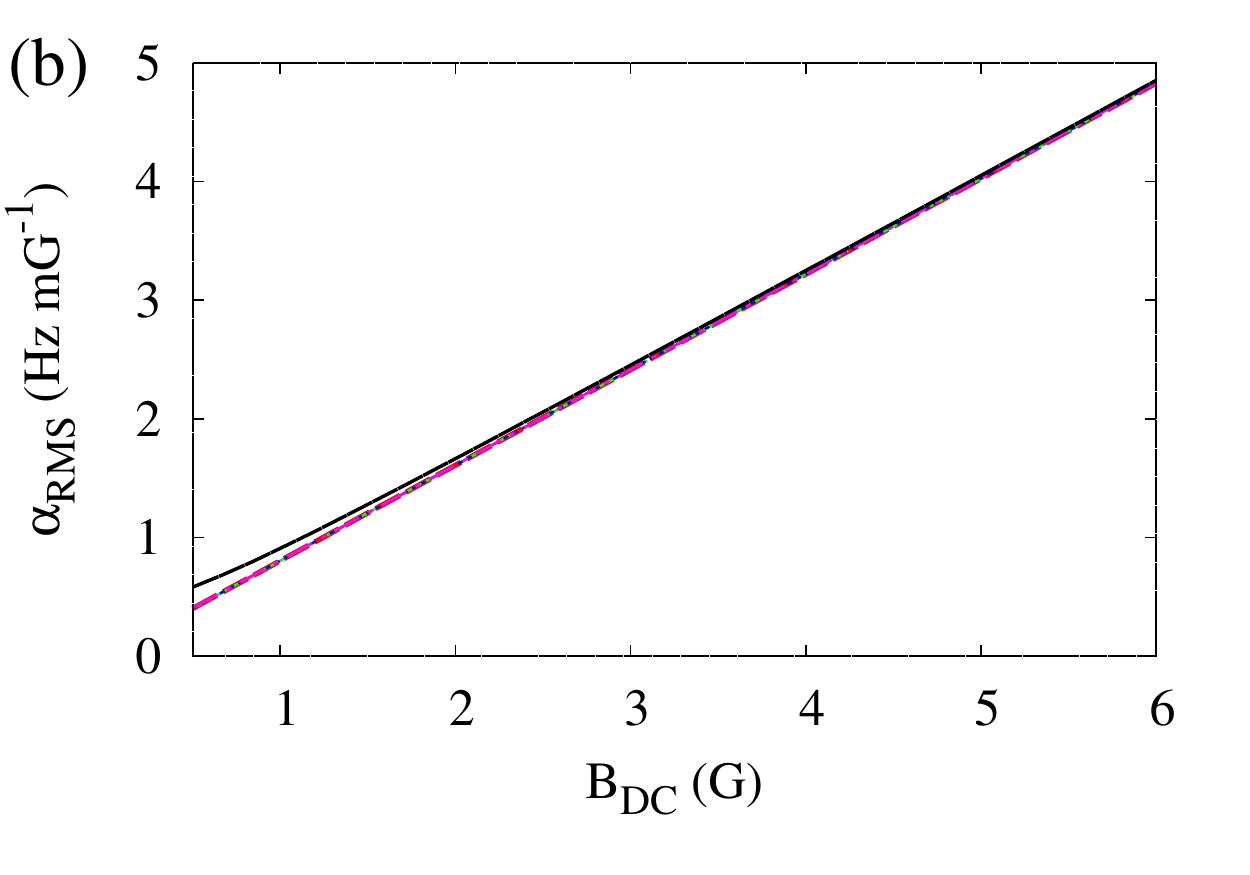}
\caption{\label{fig:RMSdifferentialmagneticmomentOPT} 
Average sensitivity $\left\langle \alpha_{\textrm{DC}} \right\rangle$ of the dressed atomic transitions with respect to the static field (Eq. \ref{eq:alphaRMS}), calculated using the field configuration optimised numerically. Panel (a) shows $\left\langle \alpha_{\textrm{DC}} \right\rangle$ as a function of the ratio $B_{\textrm{RF}}^{+}/B_{\textrm{DC}}$ for the DC fields $0.5$\,G (short-dashed, red), $1.70$\,G (dashed, green), $3.08$\,G (dotted, blue), $4.28$\,G (dot-dashed, cyan) and $6.0$\,G (solid, black). Panel (b) shows $\left\langle \alpha_{\textrm{DC}} \right\rangle$ as a function of the static field, $B_{\textrm{DC}}$, for  RF fields of amplitude $0.5$\,G (short-dashed, red), $1.70$\,G (dashed, green), $3.08$\,G (dotted, blue), $4.28$\,G (dot-dashed, cyan) and $6.0$\,G (solid, black). }
\end{figure}

When using the optimised frequencies, $\left\langle \alpha_{\textrm{DC}}\right\rangle$ is reduced by one order of magnitude and becomes weakly dependent on the amplitude of the RF field (since we compensate for their main contribution to the energy shifts). It also becomes linearly dependent on the static field, as shown in Fig. \ref{fig:RMSdifferentialmagneticmomentOPT} \footnote{When using the RWA conditions, we have $ \frac{\partial \left\langle \alpha_{\textrm{DC}} \right\rangle}{\partial B_{\textrm{DC}} } \approx 0$ and $ \frac{\partial \left\langle \alpha_{\textrm{DC}} \right\rangle}{\partial B_{\textrm{RF,+}} } > 0$. With the optimized RF frequencies $ \frac{\partial \left\langle \alpha_{\textrm{DC}} \right\rangle}{\partial B_{\textrm{DC}} } > 0$ and $ \frac{\partial \left\langle \alpha_{\textrm{DC}} \right\rangle}{\partial B_{\textrm{RF,+}} } \approx 0$.}. 

\begin{figure}[!htb]
\centering
\includegraphics[height=0.75\textwidth]{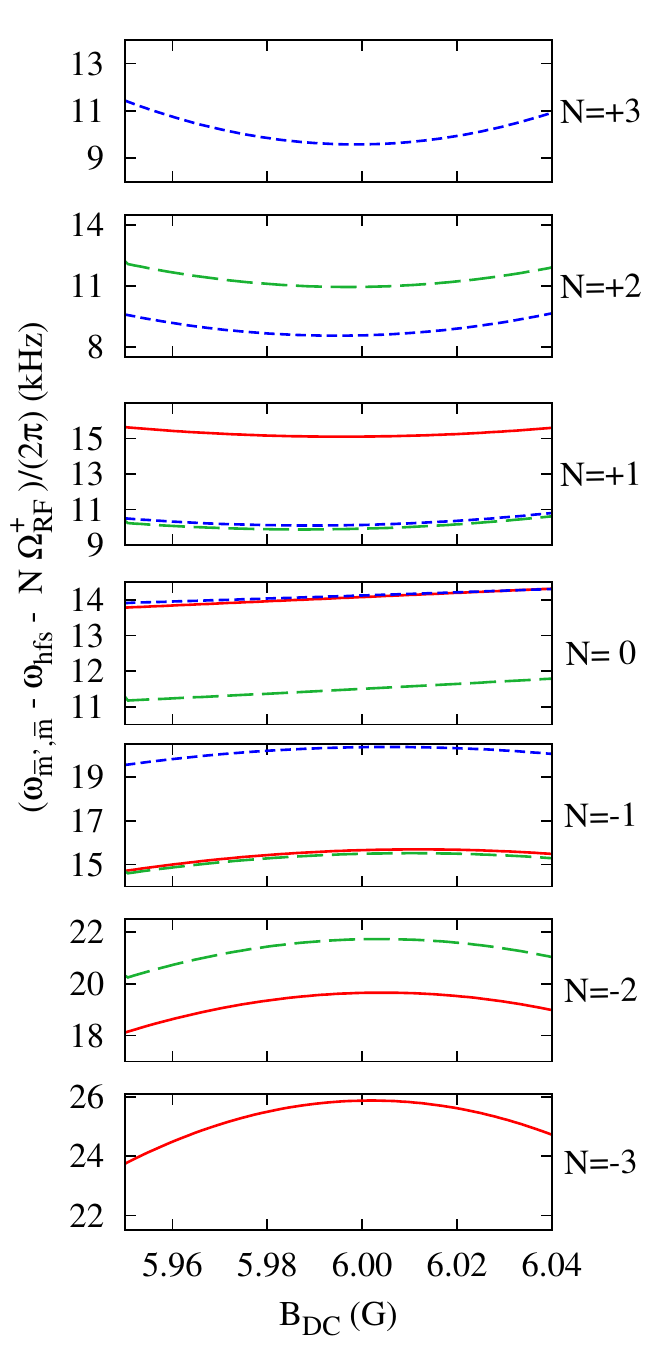}
\caption{\label{fig:threestates3} Detuning of the resonant frequencies of all hyperfine transitions in RF dressed 87Rb, as functions of the static magnetic field. The field configuration is numerically optimised to minimise the RMS of the linear sensitivity of all bichromatic dressed transitions (see text). In all cases $B_{\textrm{DC}}^0=6.0~$G and the $B_{\textrm{RF}}^+=0.12~$G. Solid, dashed and short-dashed lines correspond to transitions with the initial states $|F=1,\bar{m} \rangle=$ $|1,-1\rangle$, $|1,0\rangle$, $|1,1\rangle$, respectively. The labels on the right hand side indicate the final states as $|F=2,\bar{m}'= N - \bar{m} \rangle$.}
\end{figure}

The effect of using corrected RF frequencies is presented in Fig. \ref{fig:threestates3}, where we plot the $15$ dressed transition frequencies as functions of the static magnetic field. Note that in this case the extrema of all curves $\omega_{\bar{m}',\bar{m}}$ return to the vicinity of $B_{\textrm{DC}}=~6.0$G (compare with Figs. \ref{fig:threestates2}(a) and (b)). 

Finally, we illustrate quantitatively the improved stability enabled by the optimised bichromatic RF-dressing using the fractional frequency fluctuation due to noise in the DC field, defined by:
\begin{equation}
\sigma_{{\bar{m}',\bar{m}}} = \frac{1}{\omega_{\bar{m}',\bar{m}}} \frac{\partial \omega_{\bar{m}',\bar{m}}}{\partial B_{\textrm{DC}}} \times \Delta B_{\textrm{DC}}
\label{eq:fractionalfluctuation}
\end{equation}
For concreteness, in Table \ref{table:comparisonbaredressed}, we consider all transitions between the state $\left|1,\bar{m}=-1 \right\rangle$ to the five states of the upper hyperfine manifold, and perform this calculation for four comparable field configurations: (A) bare atom (B) monochromatic linearly polarised RF dressing with frequency $\omega_{\textrm{RF}}=\mu_{\textrm{B}}|g_2-g_1|B_\textrm{DC}/(2\hbar)$ and $B_{\textrm{RF},x} = 0.2B_{\textrm{DC}}$, (C) bichromatic dressing with the dressing field given by the RWA, Eqs. (\ref{eq:resonanceUpper})-(\ref{eq:secondOrderRatio}) and using $B_{\textrm{RF}}^{+} = 0.1B_{\textrm{DC}}$ and (D) optimised bichromatic driving with $B_{\textrm{RF}}^{+} = 0.1B_{\textrm{DC}}$. In all these cases, we consider DC fluctuations of amplitude $\Delta B_{\textrm{DC}} = 0.1~$mG. 

The best stability is obtained for the transition $\left|1,m=-1 \right\rangle \rightarrow \left|2,m=1 \right\rangle$ of the bare atom (A) at the magic field $B_{\textrm{DC}}=3.22$\,G. However, all dressing configurations considered provide a \textit{global} improvement over the fluctuations of two orders of magnitude for all transitions. In particular, the optimized bichromatic driving configuration (D) defines all transitions with the same level of protection.  

\begin{table}[!htb]
\begin{tabular}{c|c|c c c ||c}
\hline
$B_{\textrm{DC}}~$& FS & A & B & C & D \\
\hline
\hline
        & $\left|2,-2\right\rangle$ & $2.7\times10^{-7}  $ & $ 5.0\times10^{-9} $ & $ 6.8\times10^{-9} $ & $ 3.3\times10^{-11}$ \\ 
        & $\left|2,-1\right\rangle$ & $1.8\times10^{-7}  $ & $ 4.5\times10^{-9} $ & $ 4.5\times10^{-9} $ & $ 4.6\times10^{-11}$\\
$0.5~$& $\left|2,0\right\rangle$  & $9.2\times10^{-8}  $ & $ 4.0\times10^{-9} $ & $ 2.2\times10^{-9} $ & $ 6.6\times10^{-11}$\\
        & $\left|2,1\right\rangle$  & $3.1\times10^{-10} $ & $ 3.6\times10^{-9} $ & $ 2.0\times10^{-11} $ & $ 6.6\times10^{-11}$\\
        & $\left|2,2\right\rangle$  & $9.1\times10^{-8} $ & $ 3.2\times10^{-9} $ & $ 2.2\times10^{-9} $ & $ 2.3\times10^{-11}$\\
\hline

        & $\left|2,-2\right\rangle$ & $2.7\times10^{-7}  $ & $ 4.8\times10^{-9} $ & $ 6.7\times10^{-9} $ & $ 1.9\times10^{-10}$\\ 
        & $\left|2,-1\right\rangle$ & $1.8\times10^{-7}  $ & $ 4.2\times10^{-9} $ & $ 4.2\times10^{-9} $ & $ 3.5\times10^{-10}$\\
$3.2~$& $\left|2,0\right\rangle$  & $9.1\times10^{-8}  $ & $ 3.7\times10^{-9} $ & $ 1.8\times10^{-9} $ & $ 4.2\times10^{-10}$\\
        & $\left|2,1\right\rangle$  & $4.0\times10^{-12} $ & $ 3.3\times10^{-9} $ & $ 3.4\times10^{-10} $ & $ 3.7\times10^{-10}$\\
        & $\left|2,2\right\rangle$  & $9.1\times10^{-8}  $ & $ 3.0\times10^{-9} $ & $ 2.3\times10^{-9} $ & $ 1.7\times10^{-10}$\\

\hline
        & $\left|2,-2\right\rangle$ & $2.7\times10^{-7}  $ & $ 4.7\times10^{-9} $ & $ 6.5\times10^{-9} $ & $ 4.0\times10^{-10}$\\ 
        & $\left|2,-1\right\rangle$ & $1.8\times10^{-7}  $ & $ 3.8\times10^{-9} $ & $ 3.7\times10^{-9} $ & $ 7.9\times10^{-10}$\\
$7.0~$& $\left|2,0\right\rangle$  & $9.1\times10^{-8}  $ & $ 3.2\times10^{-9} $ & $ 1.3\times10^{-9} $ & $ 9.2\times10^{-10}$\\
        & $\left|2,1\right\rangle$  & $4.2\times10^{-10} $ & $ 2.9\times10^{-9} $ & $ 7.8\times10^{-10} $ & $ 7.9\times10^{-10}$\\
        & $\left|2,2\right\rangle$  & $9.1\times10^{-8}  $ & $ 2.8\times10^{-9} $ & $ 2.5\times10^{-9} $ & $ 3.8\times10^{-10}$\\
\hline
\end{tabular}
\caption{\label{table:comparisonbaredressed}
Comparison of the fractional frequency fluctuations ($\Delta\omega_{\bar{m}',\bar{m}}/\omega_{\bar{m}',\bar{m}}$) for the transitions from the state $\left|1,-1 \right\rangle$ to all final states (FS) of the $F=2$ manifold of Zeeman sub-levels (see Eq. (\ref{eq:fractionalfluctuation})). The transition frequencies and sensitivities are calculated for transitions between  (A) bare states, (B) monochromatic RF dressed states, and bichromatic RF dressed states with $\omega_{\textrm{RF}}^{\pm}$ (C) given by the RWA (Eqs. (\ref{eq:resonanceUpper})-(\ref{eq:resonanceLower})) and (D) corrected to minimise $\left\langle \alpha_{\textrm{DC}} \right\rangle$, as shown in Fig. \ref{fig:FrequenciesCorrection}. In all cases, the field fluctuations are $\Delta B =0.1~$mG. In (B) the dressing fiield is $B_{\textrm{RF}} = 0.2B_{\textrm{DC}}$, while in (C) and (D) we use $B_{\textrm{RF}}^{+} = 0.1B_{\textrm{DC}}$. In (D), $\Delta \omega_{\textrm{RF}}^{\ell}/\omega_{\textrm{RF}}^{\ell} = 2.46\times 10^{-3}~$ and $2.50\times 10^{-3}$, for $\ell=+$ and $-$, respectively. }
\end{table}

Noise in the amplitude of the dressing field  contributes to the linewidth of the spectral lines of transitions between dressed states. Temporal variations of each polar component of the RF field affect the transition frequencies  by modifying the dressed energy of the manifold it dresses and induce corrections to the perturbative energy shift of the other hyperfine manifold. We use the RMS variation of the transition frequencies ($\omega_{\bar{m},\bar{m}'}/(2\pi)$) with respect to each polar component to quantify these effects in Appendix \ref{ap:appendixRFNoise}. The average first-order sensitivity of the transition frequencies are of order $\approx 10^2~$Hz mG$^{-1}$; much larger than the residual DC sensitivity $\left\langle \alpha_{\textrm{DC}} \right\rangle \lesssim 10 ~$Hz mG$^{-1}$ achieved by tuning the frequencies of each polar component of the RF field. Thus, the improved stability of the bichromatic dressing scheme becomes useful when there is equally good stability of the RF source.  
The bi-chromatic dressing configuration has been demonstrated in \cite{Mas2019} for the case of atoms trapped in a magnetic quadrupole field. The two circular components can be produced by using two pairs of Helmholtz coils that point along the $x$ and $y$ directions, respectively. Each pair of coils is driven with an RF-signal that contains both frequencies, with the appropriate phases (for details, see \cite{Mas2019}).

\section{\label{sec:conclusions}Conclusions}

Using an ultra-cold atomic cloud of $^{87}$Rb in an optical dipole trap, we showed characteristic features of transitions between RF dressed states of the ground state hyperfine manifolds. First, considering the trio of transitions between dressed states $\left| 1,\bar{m}=-1\right\rangle \rightarrow  \{ \left| 2,\bar{m}=0 \right\rangle,\left| 2,\bar{m}=1 \right\rangle,\left| 2,\bar{m}=2 \right\rangle \} $, we observe a quadratic dependence of their transition frequencies as functions of total magnetic field $B_{\textrm{DC}}$, with a significantly weaker curvature when $\Delta \bar{m} = 0$. We found that our measurements can be explained quantitatively only after taking into account non-linear Zeeman shifts and beyond-RWA effects. However, a good qualitative description of the observed quadratic differential energy shifts can be obtained using well-known expressions for the dressed energies valid in the regimes of linear Zeeman shift and RWA.

We also study the coherence and linewidth of the transitions between RF-dressed states $\left| 1,\bar{m}=-1\right\rangle \rightarrow  \left|2,\bar{m}=1\right\rangle$ and $\left| 1,\bar{m}=-1\right\rangle \rightarrow \left|2,\bar{m}=2\right\rangle$, as functions of the applied static field. We observe a significant increase in the decay time  of Ramsey-type fringes for the transition $\left| 1,\bar{m}=-1\right\rangle \rightarrow \left|2,\bar{m}=1\right\rangle$ at a particular ``magic'' point.  Following the same method,  we observed a significant reduction of the linewidth of the transition  $\left| 1,\bar{m}=-1\right\rangle \rightarrow \left|2,\bar{m}=2\right\rangle$, reaching fluctuations of the order $\Delta \nu/\nu \approx 10^{-8}$. These experimental results demonstrate how monochromatic RF dressing can be tuned to produce pairs of selected transitions protected against DC-field noise. To further reduce the transition linewidth to the order of Hz with this scheme, the fluctuations of the amplitude of the RF fields need to be stabilised at the level of order $~10 \mu$G. Recent reports using one \cite{sinuco2019microwave}, two \cite{Mas2019}, three \cite{PhysRevA.97.013616} and four \cite{Bentine_2017} RF-frequency dressing components also find this to be a limiting factor, and indicate that active control of the RF-field amplitude is therefore necessary \cite{merkel2019magnetic}.

Furthermore, we propose a bi-chromatic RF-dressing configuration to reduce the global sensitivity of the dressed atom to noise in the static field. We demonstrate that by independently tuning the frequencies of the two circular components of the RF field, it is possible to reduce the average linear DC sensitivity to the level of Hz mG$^{-1}$. This dressing scheme enables the protection at \textit{arbitrary} DC-magnetic fields of up to $(2F+1)\times (2F'+1)=15$ atomic microwave transitions in RF-dressed $^{87}$Rb, only limited by the noise in the RF generator. Also, this bichromatic dressing configuration can stabilise more than just one single atomic transition, which is useful to define stable qdits with $d>2$. Such systems present advantages for applications in quantum metrology \cite{giovannetti2011advances} and enhanced fault-tolerance for quantum information \cite{PhysRevLett.113.230501,lanyon2009simplifying}. In addition, since the energies of the dressed states can be tuned precisely, these dressed states are attractive also for applications in quantum simulations \cite{PhysRevLett.106.190501} and for hybrid quantum systems that include devices operating in the RF/MW regime  (e.g.\ superconducting resonators, NV centres and trapped ions). Further research will be directed to determine multiparametric magic configurations in the regime of  strong RF-dressing, where multi-level atomic transitions can be made less sensitive to noise in DC and AC fields.
 
\section{Acknowledgements}
This work is supported by the project ``HELLAS-CH'' (MIS 5002735) which is implemented under the  ``Action for Strengthening Research and Innovation Infrastructures'', funded by the Operational Programme ``Competitiveness, Entrepreneurship and Innovation'' (NSRF 2014-2020) and co-financed by Greece and the European Union (European Regional Development Fund). We acknowledge financial support from the Greek Foundation for Research and Innovation (ELIDEK) in the framework of project, \emph{Guided Matter-Wave Interferometry} under grant agreement number 4823 and General Secretariat for Research and Technology (GSRT). GV received funding from the European Union’s Horizon 2020 research and innovation programme under the Marie Skłodowska-Curie Grant Agreement No  750017. This work has been supported by the UK EPSRC grant EP/M013294/1 and the University of Sussex. The authors would like to acknowledge the contribution of the COST Action CA16221.

\textbf{Authors contributions.} G.A.S-L and H.M contributed equally. H.M. and G.A.S-L. conceived and developed the main ideas of the paper. H.M. carried out the experimental measurements. H.M., S.P., G.V. and W.v.K. built the experimental setup. G.A.S.-L. and H.M. performed the data analysis. G.A.S.-L. was responsible for the theoretical and numerical work. B.M.G. and W.v.K. supervised the development of the work. All authors contributed to the result discussion and paper writing.


\appendix
\section{\label{ap:appendixA} Dressed states in the Rotating Wave Approximation}

In the main text, Eq. (\ref{eq:DressedStateFourier}) defines the unitary transformation, $U_F$, between the bare and dressed basis as:
\begin{equation}
\left\langle F,m\right. \left| \bar{F} , \bar{m} \right\rangle =  \sum_n e^{i n \omega_{\textrm{RF}}t} U_{F,m;\bar{F},\bar{m}}^n
\label{eq:UF_FourierExpansion}
\end{equation}

The Fourier coefficients of this expansion are determined by the Schr\"odinger equation and the condition Eq. (\ref{eq:FloquetState}), which translates into:
\begin{equation}
U_F^\dagger(t) \left[ H(t)  - i \hbar \partial_t \right] U_F(t) =  \sum_{\bar{F},\bar{m}}\bar{E}_{\bar{F},\bar{m}} \left| \bar{F}, \bar{m} \right\rangle \left\langle \bar{F}, \bar{m} \right|
\label{eq:FloquetStatesDef}
\end{equation}

This last expression gives us a straightforward physical interpretation of the dressed basis: the dressed states are the eigenenergy states observed in a frame of reference where the Hamiltonian is \textit{time-independent}. This concept is commonly used in quantum physics in the figure of Rotating (or Resonant) Wave Approximation (RWA), where, after moving to a rotating frame of reference the time-dependence of the Hamiltonian either cancels completely or can be neglected following perturbative arguments \cite{SERIES19781}.\\

In the present case, we consider oscillating fields with a frequency comparable to the Zeeman splitting induced by a static magnetic field, but much smaller than the hyperfine splitting ($\hbar \omega_{\textrm{RF}} \approx \mu_{\textrm{B}} g_F B_{\textrm{DC}} \ll \Delta E_{\text{Hyperfine}}$). Under these conditions, we can neglect the inter-manifold coupling and apply the RWA within each hyperfine manifold, where the transformation to the dressed basis can be written as a combination of rotations in the space of angular momentum \cite{sinuco2019microwave}:
\begin{equation}
U_F(t) = e^{i\theta_y F_y} e^{-i \frac{g_F}{|g_F|}\omega_{\textrm{RF}} t F_z} 
\end{equation}
with
\begin{equation}
\tan(\theta_y) = \frac{\sqrt{2}|\Omega_{\textrm{RF}}^{\textrm{sgn}(g_F)}|}{\omega_0 - \omega_{\textrm{RF}}}
\end{equation}
where $\hbar \omega_0 = |\mu_{\textrm{B}} g_F B_{\textrm{DC}}|$ defines the Larmor frequency and the Rabi frequency $|\Omega_{\textrm{RF}}^{\textrm{sgn}(g_F)}|$ is defined in Eq. (\ref{eq:polarcomponents}).  The corresponding dressed energies are: 
\begin{equation}
\bar{E}_{F,\bar{m}}= E_F + \textrm{sgn}(g_F) \bar{m} \sqrt{(\hbar \omega_0 - \hbar \omega_{\textrm{RF}})^2 + 2 |\hbar \Omega_{\textrm{RF}}^{\textrm{sgn}(g_F)}|^2}
\end{equation}
where $E_F = A \left(F(F+1)-I(I+1)-J(J+1)\right)/2$ is the hyperfine splitting. This dressed energy leads to the dependence of the resonant condition with respect to the field configuration:
\begin{eqnarray}
\omega_{\textrm{MW}}  &=&  n \omega_{\textrm{RF}} + \frac{(I+1/2)A}{\hbar} \nonumber \\
&& + \frac{g_{F+1}}{|g_{F+1}|} \bar{m}' \sqrt{2}|\Omega_{\textrm{RF}}^+|  - \frac{g_F}{|g_F|} \bar{m} \sqrt{2}|\Omega_{\textrm{RF}}^-| \nonumber \\ 
&&+ \left(\frac{g_{F+1}}{|g_{F+1}|} \frac{\bar{m}'}{|\Omega_{\textrm{RF}}^+|} - \frac{g_F}{|g_F|}\frac{\bar{m}}{|\Omega_{\textrm{RF}}^-|}\right)\frac{\omega_\textrm{RF}^2}{2^{3/2}} \nonumber \\  
&&- \left(\frac{\bar{m}' g_{F+1}}{|\Omega_{\textrm{RF}}^+|} - \frac{\bar{m}g_{F}}{|\Omega_{\textrm{RF}}^-|}\right) \omega_\textrm{RF}\frac{\mu_{\textrm{B}} B_{\textrm{DC}}}{\sqrt{2}\hbar}  \nonumber \\
&& + \frac{1}{2^{3/2}}\left[ \frac{g_{F+1}}{|g_{F+1}|}\frac{\bar{m}' g_{F+1}^2}{|\Omega_{\textrm{RF}}^+|} -  \frac{g_F}{|g_F|}\frac{\bar{m} g_{F}^2}{|\Omega_{\textrm{RF}}^-|} \right]\left(\frac{\mu_{\textrm{B}} B_{\textrm{DC}}}{\hbar}\right)^2  \nonumber \\
\end{eqnarray}
which is valid for inter-manifold transitions,$\left|F,\bar{m} \right\rangle \rightarrow \left| F+1,\bar{m}' \right\rangle$ and in the vicinity of the condition of resonant RF dressing, $\mu_{\textrm{B}} |g_F B_{\textrm{DC}}| \approx \hbar \omega_{\textrm{RF}}$. 

This formulation can be easily extended to situations with polychromatic driving. The time-evolution operator should be then expressed as a multidimensional Fourier series with as many dimensions as the number of applied fields with incommensurately frequencies. More explicitly, the dressed state defined in Eq. (\ref{eq:UF_FourierExpansion}) becomes:
\begin{equation}
\left\langle F,m\right. \left| \bar{F} , \bar{m} \right\rangle =  \sum_{\vec{n}} e^{i \vec{n} \vec{\omega}t} U_{F,m;\bar{F},\bar{m}}^{\vec{n}}
\label{eq:UF_PolyFourierExpansion}
\end{equation}
with $\vec{\omega}= (\omega_1, \omega_2, \ldots )$ as the vector formed with all the applied frequencies and $\vec{n}$ a vector with integer components. After plugging this Ansatz in the Schr\"odinger equation Eq. (\ref{eq:FloquetStatesDef}) we obtain a standard eigen problem defining the coefficients $U_{F,m;\bar{F},\bar{m}}^{\vec{n}}$ and generalized dressed energies $\bar{E}_{\bar{F},\bar{m}}$.

\section{\label{ap:appendixC} Microwave spectroscopy of the dressed transition $\left|1, -1\right\rangle \rightarrow \left|2, 1\right\rangle$ near the two-photon condition}

\begin{figure}[!htb]
\includegraphics[width=0.43\textwidth]{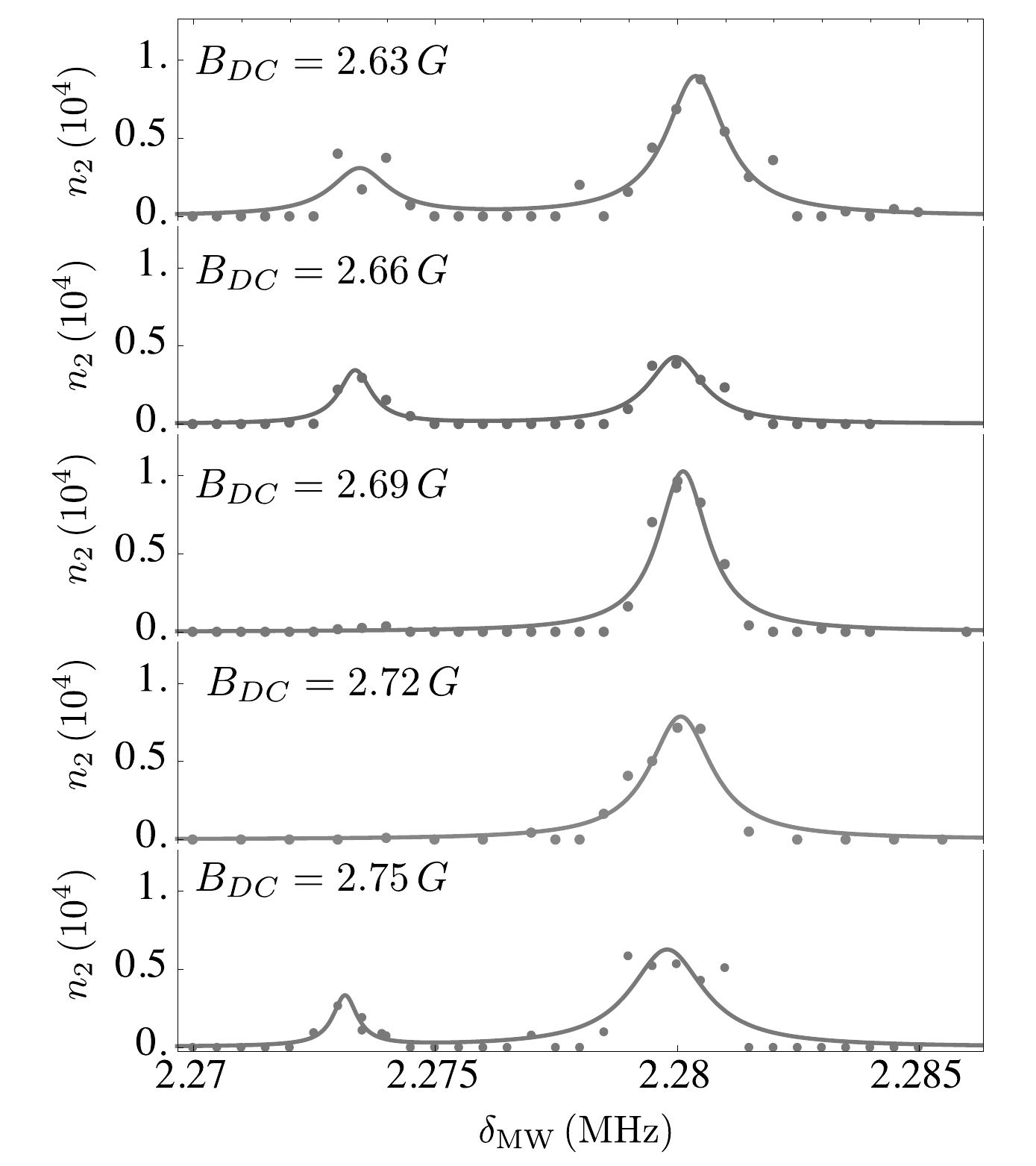}
\caption{\label{fig:peakdoubling} Spectra for selected values of DC field in Fig.~\ref{fig:ramsey2dcfields}(a). From top to bottom, $B_{\textrm{DC}}=2.63$\,G, $2.66$\,G, $2.69$\,G, $2.72$\,G and $2.75$\,G. The Rabi frequency is $\Omega_{\textrm{RF}}/2\pi$\,kHz and the RF frequency is $\omega_{\textrm{RF}}/2\pi=2.27$\,MHz. We observe the appearance of a second, usually weaker transition in some cases. Just like in the rest of the text, $n_2$ is the atom number measured with absorption imaging in $F=2$. Each point corresponds to one measurement. We fit Lorentzian curves for each observed peak. We observe the intermittent appearance of a second, weaker peak. We attribute this second peak to transitions from atoms that did not follow adiabatically our dressing sequence.}
\end{figure}

We measured the line-shift of the transition $|1,\bar{m}=-1\rangle \rightarrow |2,\bar{m}=1\rangle$ in Fig.~\ref{fig:ramsey2dcfields}(a). We intermittently observed a second weaker peak, with some examples in Fig.~\ref{fig:peakdoubling}. Here, we interrogate the first group of transitions, i.e. those with $N=1$, following Eq. (\ref{eq:resonantcondition}), with $\omegaRF/2\pi=2.27$\,MHz. A qualitative analysis indicates that the each peak is produced by atoms in different dressed states, with the intensity of each corresponding to their population. This second resonance indicates that some atoms do not follow adiabatically our dressing sequence \cite{PhysRevA.96.023429}.

\section{\label{ap:appendixRFNoise} Off-resonant corrections to the RWA and linear sensitivity of the bichromatic driven atom to RF amplitude noise}

The susceptibility of the dressed transitions with respect to variations of the dressing parameters can be calculated correcting the RWA dressed energy by including perturbative shifts of the Zeeman states: 
\begin{eqnarray}
\bar{E}_{F,\bar{m}}&=&  E_F +\frac{g_F}{|g_{F}|} \bar{m} \times \nonumber \\
&&\sqrt{(\mu_{\textrm{B}} |g_F| B_{\textrm{DC}} - \hbar \omega_{\textrm{F}} + \hbar \Delta_{F})^2 +  2\hbar |\Omega_{\textrm{RF}}^{\textrm{sgn}(g_F)}|^2} \nonumber
\label{eq:LZ_RWA_dressedenergies}
\end{eqnarray}
with $E_F=A(F(F+1)-I(I+1)-J(J+1))/2$ and where $\Delta_F$ has contributions from the fields which are counter-rotating in the dressed frame of reference. The total shift can be approximated by:

\begin{eqnarray}
\hbar \Delta_{F}  &=&   \frac{1}{2} \frac{g_{F}}{|g_{F}|}\frac{|\mu_{\textrm{B}} g_{F} (B^{-\textrm{sgn}(g_F)}_{\textrm{RF}} + \Delta B_{\textrm{RF}}^{\textrm{sgn}(g_F)})|^2}{\mu_{\textrm{B}} |g_{F}|B_{\textrm{DC}} + \hbar \omega_0} 
\label{eq:minusCorrection}
\end{eqnarray}
where $\omega_0 = (\omega_+ + \omega_-)/2$ and $\Delta B_{\textrm{RF}}^\ell$ is the counter-rotating component of the field oscillating at frequency $\omega_{\ell}$.

Another important contribution to the broadening of the transition lines is their instability with respect to variations in the amplitude of the dressing fields. In this case, we can distinguish four contributions emerging from the decomposition of the variations of each dressing frequency into $\sigma_+$ and $\sigma_-$ polarisations. To evaluate the effects of fluctuations of the RF field, we split the noise of each polar component of the RF field into co-rotating and counter-rotating contributions:
\begin{equation}
B_{\textrm{RF}}^{\ell} = B_{\textrm{RF},0}^{\ell} + \delta B_{\textrm{RF},\ell}^{\ell} +\delta B_{\textrm{RF},-\ell}^{\ell}
\end{equation}
where $\ell \in {+,-}$, $B_{\textrm{RF},0}^{\ell}$ is the RMS value of the field, and  $\delta B_{\textrm{RF},(-)\ell}^{\ell}$ is the component of the fluctuation co-rotating (counter-rotating) with the $\sigma_\ell$ ($\sigma_{-\ell}$) component of the RF field. 
As in Sec. \ref{sec:sensitivity}, we quantify these effects by defining average linear sensitivities, $\left\langle \alpha_{\textrm{RF},m}^{\ell} \right\rangle$ for each component of the ($m\in{\{+,-\}}$) of the two dressing fields $\ell \in{\{+,-\}}$:
\begin{equation}
\left\langle\alpha_{\textrm{RF},\ell}^{m}\right\rangle  = \sqrt{ \frac{1}{15} \sum_{m=-1}^1\sum_{m'=-2}^2 \left(\frac{\partial \omega_{\bar{m},\bar{m}'}} {\partial \delta B_{\textrm{RF},m}^{\ell}} \right)^2}
\end{equation}

\begin{figure}[!htb]
\centering
\includegraphics[width=0.48\textwidth]{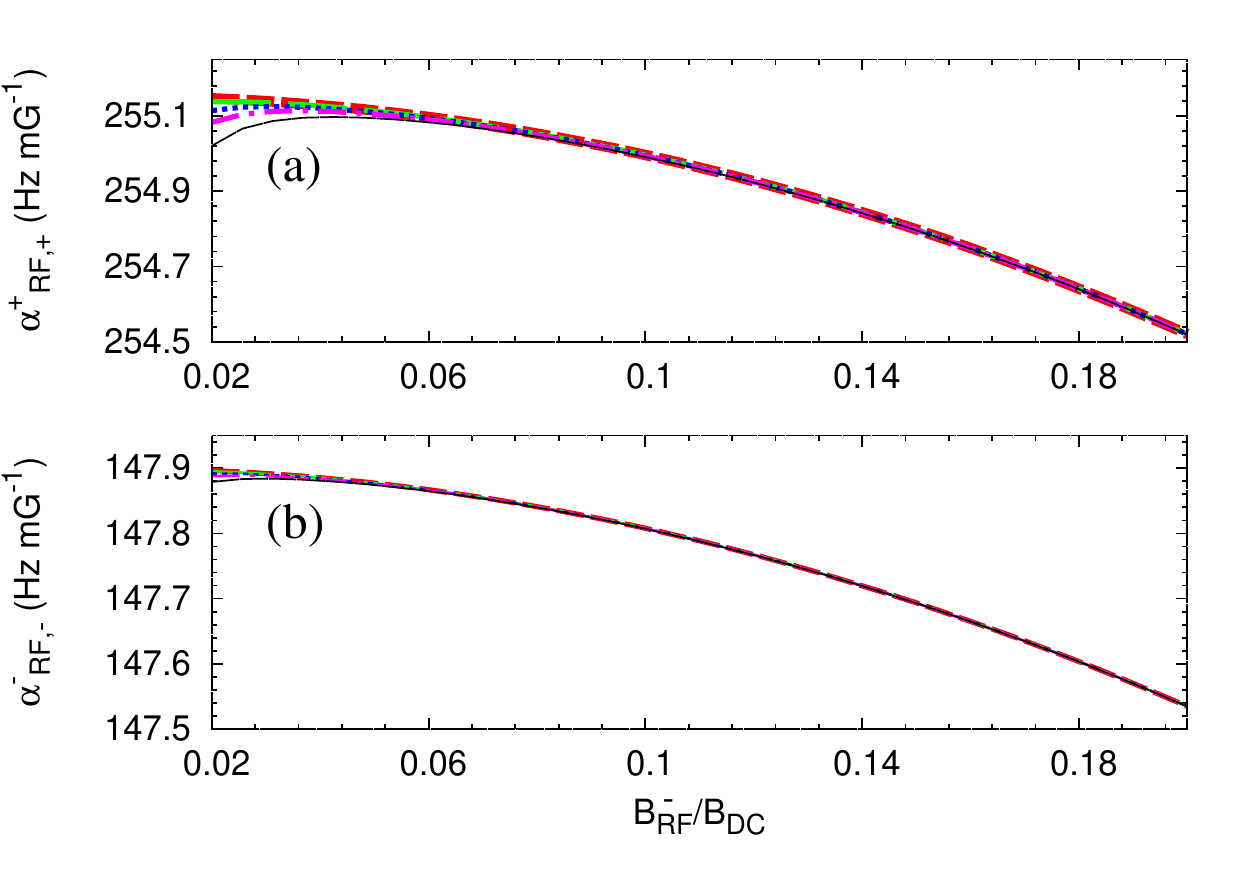}
\caption{\label{fig:RFsensitivityA} RMS susceptibility of the dressed transition frequencies of $^{87}$Rb with respect to co-rotating  variations of the (a) $\sigma_{+}$ and  (b)  $\sigma_{-}$ dressing fields as functions of the dressing field amplitude. The static field are $0.5~$G  (dashed line), $1.5~$G (long-dashed line), $3.0~$G (dotted line), $4.5~$G(dashed-dotted line) and $6.0~$G (solid line). For each dressing configuration, we evaluate the optimal combination of $\sigma_{\pm}$ frequencies that minimise $\left\langle \alpha_{\textrm{DC}}\right\rangle$ Eq. (\ref{eq:alphaRMS})}
\end{figure}

In Fig. \ref{fig:RFsensitivityA} we show the atomic sensitivity associated with the co-rotating noisy components of the dressing fields ($m=\ell$), when using the optimised frequency configuration. The co-rotating noise modifies the amplitude of the dressing field, leading to a sensitivity that weakly depends on the static field. 

\begin{figure}[!htb]
\centering
\includegraphics[width=0.48\textwidth]{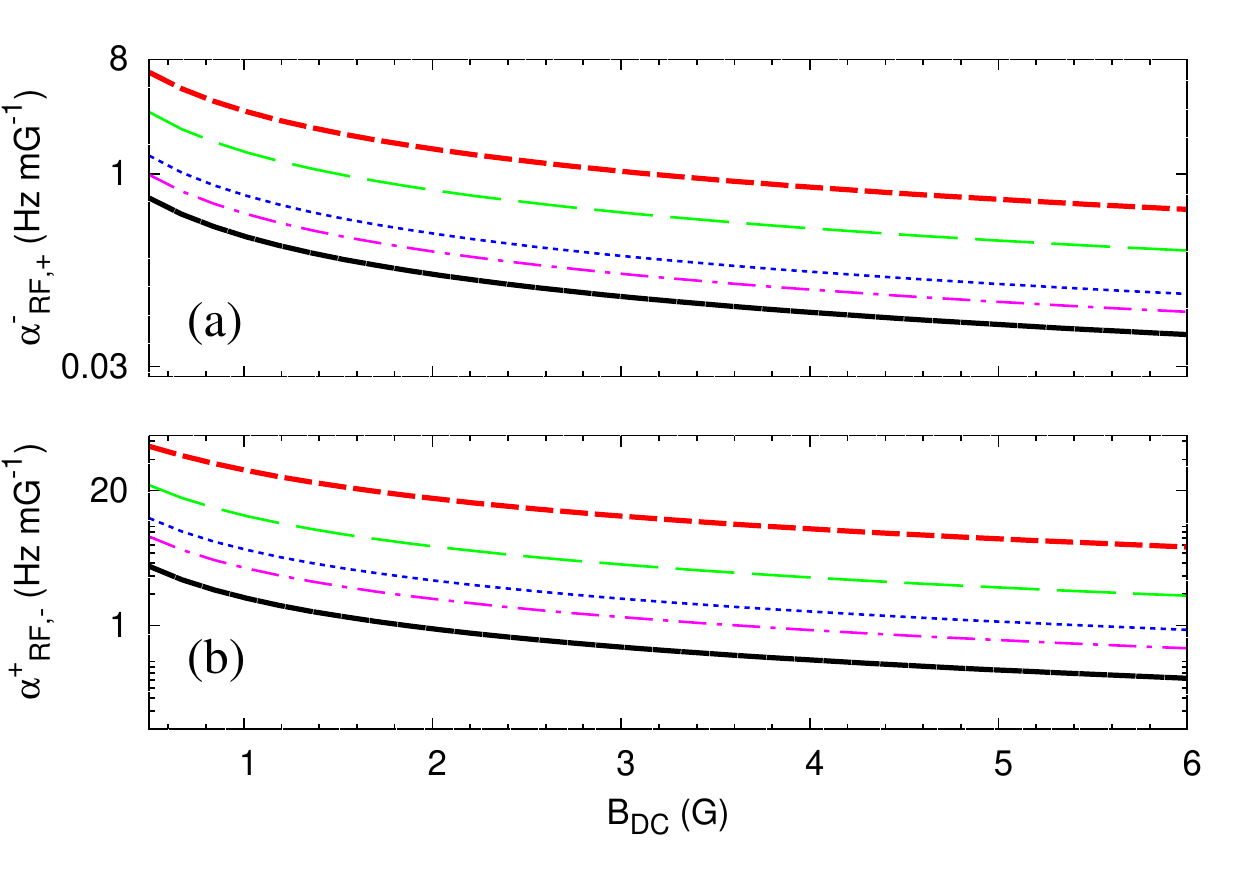}
\caption{\label{fig:RFsensitivityB} RMS linear sensitivity of the dressed transition frequencies of $^{87}$Rb with respect to the counter-rotating polar components (a) $\sigma_{+}$ (b) $\sigma_{-}$ of the variations of the dressing fields $\sigma_{-}$ (a) and $\sigma_{+}$ (b). The scaled amplitudes of the dressing fields are $0.02~B_{\textrm{DC}}$ (dashed line), $0.03~B_{\textrm{DC}}$ (long-dashed line), $0.1~B_{\textrm{DC}}$ (dotted line), $0.15~B_{\textrm{DC}}$ (dashed-dotted line) and $0.2~B_{\textrm{DC}}$ (solid line). For each dressing configuration, we evaluate the optimal combination of $\sigma_{\pm}$ frequencies that minimise $\left\langle \alpha_{\textrm{DC}}\right\rangle$.}
\end{figure}

Another effect due to noise of the dressing field comes from the counter-rotating noisy component, which causes off-resonant energy shifts similar to the ones described before in Eq (\ref{eq:minusCorrection}), but this time oscillating at the same frequency of the corresponding dressing component. In  Fig. \ref{fig:RFsensitivityB} we show the a RMS linear susceptibility of all transitions, $\left\langle \alpha_{\textrm{RF},-\ell}^{\ell}\right\rangle$, induced by counter-rotating variations of the dressing fields, using the frequency configuration that minimise the sensitivity to static fields.

The scaling and order of magnitude of $\left\langle \alpha_{\textrm{RF},\ell}^{m}\right\rangle$ can be obtained by calculating second order perturbative energy shifts and applying the RWA (see Appendix \ref{ap:appendixC}). Note that the sensitivity to variations of the dressing fields is two orders of magnitude larger than the sensitivity to variations of the static field with the optimised parameters (compare with  Fig. \ref{fig:RMSdifferentialmagneticmomentOPT}). To reduce the line broadening of resonant transitions to the level of a few Hz, the fluctuations of dressing RF fields should of order $\approx 10~\mu$G, which corresponds to a relative amplitude fluctuation of the order $10^{-5} - 10^{-6}$.

\bibliographystyle{apsrev4-1}
\bibliography{ProtectionReferences}

\end{document}